\documentclass[prd,aps,showpacs,nofootinbib,showkeywords,eqsecnum,twocolumn]{revtex4-1}

\usepackage{graphicx,color,amsmath,amsxtra}
\usepackage{epsf}
\usepackage{amssymb}
\usepackage{enumerate}
\usepackage{hhline}
\usepackage{array}
\usepackage{tabularx}
\usepackage{hangcaption}
\usepackage[unicode]{hyperref}
\usepackage{graphicx}              
\usepackage{epstopdf}

\hypersetup{
  colorlinks   = true, 
  urlcolor     = blue, 
  linkcolor    = blue, 
  citecolor   = blue 
}

\begin{document}
\pagestyle{plain}
\pagestyle{myheadings}

\title{Higher dimensional massive bigravity}
\author{Tuan Q. Do}
\email{tuanqdo@vnu.edu.vn} \email{tuanqdo.py97g@nctu.edu.tw}
\affiliation{Faculty of Physics, VNU University of Science, Vietnam National University, Hanoi 120000, Vietnam}
\date{\today }

\begin{abstract}
We study higher dimensional scenarios of massive bigravity, which is a very interesting extension of nonlinear massive gravity since its reference metric is assumed to be full dynamical.  In particular, the Einstein field equations along with the following constraint equations for both physical and reference metrics of a five-dimensional massive bigravity will be addressed. Then, we study some well-known cosmological spacetimes such as the Friedmann-Lemaitre-Robertson-Walker, Bianchi type I, and Schwarzschild-Tangherlini metrics  for the five-dimensional massive bigravity. As a result, we find that massive graviton terms will serve as effective cosmological constants in both physical and reference sectors if a special scenario, in which reference metrics are chosen to be proportional to physical ones, is considered for all mentioned metrics. Thanks to the constancy property of massive graviton terms, consistent cosmological solutions will be figured out accordingly. 

\end{abstract}

%


\maketitle
\section{Introduction} \label{sec1}
Recently, a nonlinear massive gravity has been successfully constructed by de Rham, Gabadadze, and Tolley (dRGT)  \cite{RGT} as a generalization of massive gravity proposed by Fierz and Pauli in a seminal paper \cite{FP,NAH}. As a result, the most important property of the dRGT theory is that it has been proved  to be free of the so-called Boulware-Deser (BD) ghost \cite{BD} by different approaches  whatever the form of reference (or  fiducial) metric \cite{proof}.  In fact, many cosmological and physical aspects  of the dRGT theory have been investigated extensively, which can be found in recent interesting review papers \cite{review}.

It is known that  a second metric $f_{ab}$ in the massive gravity, which is usually called a reference (or fiducial) metric to distinguish it from  a dynamical physical metric $g_{\mu\nu}$, is assumed to be non-dynamical. It is introduced along with the St\"uckelberg scalar fields $\phi^a$ to give a manifestly diffeomorphism invariant description \cite{NAH}. Since the dRGT theory has been proved to be free of BD ghost for arbitrary reference metric \cite{proof}, one can therefore extend this theory to a more general scenario, in which the reference metric can be dynamical.  Note that a theory involving both dynamical metrics has been known as a bimetric gravity (or bigravity for short) theory \cite{Isham}. Similar to the massive gravity, the old bigravity \cite{Isham} has  faced  the same BD ghost problem for quite long time.  However, after the discovery de Rham, Gabadadze, and Tolley, a  ghost-free nonlinear bigravity employing the massive graviton terms (or the interaction terms) of nonlinear massive gravity has been proposed by Hassan and Rosen in {Ref.~}\cite{SFH}. As a result, there are two  gravitons interacting with each other in the ghost-free bigravity theory, one is massive carrying five degrees of freedom  and another is massless carrying two degrees of freedom.  Soon after this investigation, ghost-free multi-metric gravity (or multi-gravity for short) theories have also been formulated in {Refs.~}\cite{NK,Hassan:2012wr}. Of course, the number of gravitons in multi-gravity must be larger than two. However, the number of massive gravitons is always larger than the number of massless gravitons in multi-gravity, which should be equal to one.    

It turns out that the massive bigravity has  received a lot of discussions recently. For a up-to-date  review on the progress of the bigravity, see {Ref.~}\cite{review-bigravity}.  In particular, the bigravity has been discussed extensively in {Refs.~}\cite{MS,MSV,MSV1,aoki,blackholes,blackholes-stability,blackholes-review,wormhole,soda,mod1,mod2,non-minimal,higher-bigravity,higher-bigravity-more}. More precisely, some cosmological issues of the ghost-free bigravity such as  the cosmological evolution and the dark matter problem have been examined in {Refs.~}\cite{MS,MSV1,aoki}; while some black holes, wormholes, and some anisotropic Bianchi types  have been investigated in the context of bigravity in Refs. \cite{blackholes,blackholes-stability,blackholes-review}, {Ref.~}\cite{wormhole}, and {Ref.~}\cite{soda}, respectively. For recent reviews on the black holes solutions of massive (bi)gravity, see Ref.~\cite{blackholes-review}. Furthermore, some extensions of the massive bigravity have been proposed, e.g., the $f(R)$ bigravity in Ref.~\cite{mod1}, the scalar-tensor bigravity in Ref.~\cite{mod2}, and the massive bigravity with non-minimal coupling of matter in Ref. \cite{non-minimal}. Along this line, another natural way to generalize the bigravity is constructing higher dimensional scenarios of massive bigravity as done in papers listed in {Refs.~}\cite{higher-bigravity,higher-bigravity-more}. However, these paper have not discussed  particularly the well-known Friedmann-Lemaitre-Robertson-Walker, Bianchi type I, and Schwarzschild-Tangherlini metrics with additional higher dimensional massive graviton terms, which must vanish in all four-dimensional spacetimes but do exist in any higher dimensional spacetime. As far as we know, many previous papers have focused only, even when they discuss higher dimensional solutions of massive (bi)gravity, on  the first three graviton terms ${\cal U}_i$'s (or ${\cal L}_i$'s/2) ($i=2-4$), which have been shown to be non-vanishing in any four-dimensional spacetime \cite{higher-models}. Therefore, it is physically important to study higher-than-four dimensional massive (bi)gravity involving not only the first three graviton terms, ${\cal U}_2$, ${\cal U}_3$, and ${\cal U}_4$, but also ${\cal U}_{i>4}$'s terms since this inclusion might affect on the previous results ~\cite{higher-models} due to the existence of additional graviton terms ${\cal U}_{i>4}$, which would not vanish in higher dimensional spacetimes.

It is noted that we have been able to construct explicit ghost-free higher dimensional graviton terms such as ${\cal L}_5$, ${\cal L}_6$, and ${\cal L}_7$ in five-, six-, and seven-dimensional spacetimes, respectively, in recent works on the higher dimensional nonlinear massive gravity \cite{review-bigravity,higher-bigravity,higher-bigravity-more,TQD}.  It is apparent that our construction is based on the well-known Cayley-Hamilton (CH) theorem in linear algebra for the determinant of square matrix \cite{CH-theorem}. For detailed discussions on how to construct all graviton terms based on the CH theorem, see for example {Refs.~}\cite{review-bigravity,TQD}. Note that this method can be used to build up any higher dimensional graviton term  for both dRGT gravity and massive  bigravity theories.  Additionally, we have also shown in {Ref.~}\cite{TQD} that some results obtained in a four-dimensional nonlinear massive gravity \cite{WFK} will be recovered in a five-dimensional scenario by fine-tuning $\alpha_5$, the coefficient of additional graviton term ${\cal L}_5$ in the action, such that a specific relation between $\alpha_5$ and the other coefficients $\alpha_3$ and $\alpha_4$ is satisfied. 

As a companion paper to {Ref.~}\cite{TQD} and {Refs.~}\cite{higher-bigravity,higher-bigravity-more}, the present paper is devoted to study a five-dimensional scenario of massive bigravity with an additional graviton term, ${\cal U}_5={\cal L}_5/2$, which disappears in all four-dimensional spacetimes but survives in any higher-than-four dimensional one. In particular, we will examine whether the graviton terms ${\cal U}_i$'s ($i=2-5$) act as  effective cosmological constants in a number of spacetimes such as the Friedmann-Lemaitre-Robertson-Walker (FLRW), Bianchi type I, and Schwarzschild-Tangherlini for both physical and reference metrics, which will be assumed to be compatible with each other. It is noted that the reference metric $f_{\mu\nu}$ in the bigravity is dynamical, similar to the physical metric $g_{\mu\nu}$, rather than non-dynamical. Hence, the field equations of $f_{\mu\nu}$ in the massive bigravity will be differential rather than algebraic as in the dRGT gravity. Hence, the graviton terms could not easily turn out  to be effective constants. However, the Bianchi identity will be applied to the reference metric since its role in the action is now similar to that of the physical metric. As a result, the Bianchi constraints of $f_{\mu\nu}$ along with that of $g_{\mu\nu}$ will lead to some solutions, under which the graviton terms could act as effective cosmological constants in both physical and reference sectors (from now on we will call them the $g$- and $f$-sectors for short). Consequently, the field equations in both of these sectors could become simple to be solvable analytically (or numerically). Indeed, we will figure out some simple solutions of the five-dimensional massive bigravity with effective cosmological constants coming from the graviton terms ${\cal U}_i$'s ($i=2-5$) for all above mentioned metrics. We will also discuss whether the five-dimensional bigravity recovers results of the four-dimensional bigravity.

This paper will be organized as follows. A brief introduction of this research has been given in section \ref{sec1}. Some basic details of four-dimensional massive  bigravity will be presented in section \ref{sec2}. A five-dimensional massive bigravity model will be shown in section \ref{sec3}. The FLRW, Bianchi type I, and  Schwarzschild-Tangherlini metrics will be studied in the framework of the five-dimensional bigravity in section \ref{sec4}, section \ref{sec5}, and section \ref{sec6}, respectively. In section \ref{sec7}, we will examine whether effective cosmological constants derived from the graviton terms of  four-dimensional massive bigravity will be recovered in the context of five-dimensional massive bigravity. Finally, concluding remarks and discussions will be given in section \ref{con}.  
\section{Four-dimensional massive bigravity} \label{sec2}
In this section, we will brieftly review the four-dimensional massive bigravity \cite{SFH}, which is based on the four-dimensional nonlinear massive gravity \cite{RGT,review}. In particular, an action of the four-dimensional massive bigravity is given by \cite{SFH}
\begin{eqnarray} \label{action1}
S_{\text {4d}}&=& M_g^2 \int {d^4 } x\sqrt { g} R(g)+ M_f^2 \int {d^4 } x\sqrt {f} R(f) \nonumber\\
&&+2m^2 M_{\text{eff}}^2 \int {d^4 } x\sqrt { g} \Bigl( {\cal U}_2 +\alpha_3 {\cal U}_3 +\alpha_4 {\cal U}_4 \Bigr),\nonumber\\
\end{eqnarray}
where $g\equiv -\det g_{\mu\nu}$, $f\equiv -\det f_{\mu\nu}$, and ${\cal U}_i$'s $(i=2-4)$ are massive graviton terms (or interaction terms) defined in terms of ${\cal K}^\mu_\nu \equiv {\delta}^\mu_\nu- \sqrt{g^{\mu\sigma}f_{\sigma\nu}}$ as follows
\begin{eqnarray}
{\cal U}_2 &=& \frac{1}{2}\Bigl\{ [{\cal K}]^2-[{\cal K}^2]\Bigr\}, \\
{\cal U}_3 &=& \frac{1}{6} \Bigl\{ [{\cal K}]^3-3[{\cal K}][{\cal K}^2]+2[{\cal K}^3]\Bigr\}, \\
{\cal U}_4 &=& \frac{1}{24}\Bigl\{ [{\cal K}]^4-6[{\cal K}]^2 [{\cal K}^2]+3[{\cal K}^2]^2+8 [{\cal K}][{\cal K}^3] -6[{\cal K}^4]\Bigr\}. \nonumber\\
\end{eqnarray}
 It is noted that the other parameters, $\alpha_0$ associated with ${\cal U}_0 = 1$ and $\alpha_1$ associated with ${\cal U}_1 = [{\cal K}]$, have been chosen to be zero since we would like to have flat space solutions for the corresponding field equations in the weak field limit, $g_{\mu\nu} \approx f_{\mu\nu} \approx \eta_{\mu\nu}$. On the other hand, $\alpha_2$ associated with ${\cal U}_2$ has  also been set to be one in order to recover  the Fierz-Pauli term  \cite{SFH}. Hence, we end up with two free parameters $\alpha_3$ and $\alpha_4$. It is also noted that one can work in an equivalent framework constructed by Hassan and Rosen \cite{SFH}, where the graviton terms ${\cal U}_n $'s will no longer be functions of ${\cal K}^\mu_\nu$ but of ${\mathbb X}^\mu_\nu \equiv \sqrt{g^{\mu\sigma}f_{\sigma\nu}}$, while the parameters $\alpha_i$'s will be replaced by $\beta_i$'s. Furthermore, the highest order of graviton terms in the Hassan-Rosen framework is third order rather than fourth order as in the dRGT framework.  Hence, working in the Hassan-Rosen framework might be more convenient than the dRGT framework. For more details, especially the relation between the coefficients $\alpha_i$'s and $\beta_i$'s, see \cite{SFH}. In this paper, however, we prefer using the original definitions of the dRGT massive gravity ~\cite{RGT} in order to compare obtained results in this paper with that investigated in the previous paper on the dRGT theory \cite{TQD}.

It is noted  in the above action that  the square brackets stand for the trace of matrix, i.e., $[{\cal K}] = \text{tr} {\cal K}$, $[{\cal K}^n] = \text{tr} {\cal K}^n$, and $[{\cal K}]^n = (\text{tr} {\cal K})^n$. Note that St\"uckelberg scalar fields  will  no longer be introduced in the context of bigravity. 
In addition, $m$ is the graviton mass, while $R(g)$  and $R(f)$ stand for the scalar curvatures of a physical metric $g_{\mu\nu}$ and a reference  metric $f_{\mu\nu}$, respectively. In the rest of paper, we will still use the name "reference metric" for the $f_{\mu\nu}$ in order to distinguish it from the "physical metric" $g_{\mu\nu}$.  Of course, one can rename $f_{\mu\nu}$ as the "second metric" since $f_{\mu\nu}$ plays a similar role as $g_{\mu\nu}$ does in the context of massive bigravity.  In addition,  $M_{\text{eff}}$ is an effective Planck mass defined in terms of two other Planck masses, $M_g$ for the physical metric $g_{\mu\nu}$ and $M_f$ for the reference metric $f_{\mu\nu}$, as follows \cite{SFH}
\begin{equation}
M_{\text{eff}}^2 = \left(\frac{1}{M_g^2}+ \frac{1}{M_f^2}\right)^{-1}.
\end{equation}
It is noted that the  reference metric $f_{\mu\nu}$ in the bimetric gravity has been regarded as a full dynamical metric as the physical one \cite{SFH}, while the reference  metric in the nonlinear massive gravity theory has been chosen to be non-dynamical \cite{RGT}.  This is a main difference between these two theories. As a result, choosing $f_{\mu\nu}$ as a full dynamical metric will yield a theory invariant under general coordinate transformations without introducing  the St\"uckelberg scalar fields \cite{SFH}. Additionally, the indexes of tensors  in the $f$-sector of bigravity will be raised or lowered by the reference metric instead of the physical metric since the reference metric is chosen to play the same role as the physical metric \cite{SFH}.

As a result, the Einstein field equations for the  $g$-sector associated with the physical metric $g_{\mu\nu}$ can be derived to be \cite{SFH}
\begin{equation}
M_g^2\left(R_{\mu\nu}(g)-\frac{1}{2}g_{\mu\nu}R(g)\right) + m^2 M_{\text{eff}}^2 {\cal H}_{\mu\nu}(g) =T_{\mu\nu}(g),
\end{equation}
where
\begin{equation}
{\cal H}_{\mu\nu}(g) = X_{\mu\nu}(g) + \alpha_4 Y_{\mu\nu}(g) ,
\end{equation}
\begin{eqnarray} \label{eqX}
X_{\mu\nu}(g) &=& {\cal K}_{\mu\nu} -[{\cal K}]g_{\mu\nu} \nonumber\\
&& -(\alpha_3+1) \Bigl\{{{\cal K}_{\mu\nu}^2-[{\cal K}]{\cal K}_{\mu\nu}+ {\cal U}_2  g_{\mu\nu} }\Bigr\} \nonumber\\
&& +(\alpha_3+\alpha_4) \Bigl\{{{\cal K}_{\mu\nu}^3-[{\cal K}]{\cal K}_{\mu\nu}^2+{\cal U}_2  {\cal K}_{\mu\nu} }\Bigr\} \nonumber\\
&& - \left(\alpha_3 +\alpha_4 \right) {\cal U}_3 g_{\mu\nu},\\
 \label{eqY}
Y_{\mu\nu}(g) &=& -{\cal U}_4 g_{\mu\nu} + \tilde Y_{\mu\nu}(g),\\
\label{eq-hatY}
\tilde Y_{\mu\nu}(g) &=& {\cal U}_3 {\cal K}_{\mu\nu} - {\cal U}_2 {\cal K}^2_{\mu\nu}
+[{\cal K}]{\cal K}^3_{\mu\nu} -{\cal K}^4_{\mu\nu}.
\end{eqnarray}
Is is straightforward to show that the tensor $Y_{\mu\nu}(g)$ always vanishes \cite{SFH,WFK}. On the other hand, the field equations for the $f$-sector associated with the reference metric $f_{\mu\nu}$ turn out to be \cite{SFH}
\begin{equation} \label{equation-f}
\sqrt {f} M_f^2\left(R_{\mu\nu}(f)-\frac{1}{2}f_{\mu\nu}R(f)\right)+ \sqrt{g} m^2 M_{\text{eff}}^2 s_{\mu\nu}(f) =0,
\end{equation}
where 
\begin{eqnarray} \label{eq-s}
s_{\mu\nu}(f) &=& - {\cal K}^{\sigma}_{\mu} f_{\sigma \nu} + \left\{ [{\cal K}]+\alpha_3 {\cal U}_2  +\alpha_4 {\cal U}_3 \right\} f_{\mu\nu}\nonumber\\
&&+(\alpha_3+1) \Bigl\{{ {\cal K}^{\rho}_{\mu}{\cal K}^{\sigma}_{\rho}-[{\cal K}]{\cal K}^{\sigma}_{\mu}}\Bigr\} f_{\sigma\nu} \nonumber\\
&& -(\alpha_3+\alpha_4) \Bigl\{{ {\cal K}^{\rho}_{\mu}{\cal K}^{\delta}_{\rho}{\cal K}^{\sigma}_{\delta} -[{\cal K}]{\cal K}^{\rho}_{\mu}{\cal K}^{\sigma}_{\rho}+{\cal U}_2 {\cal K}^{\sigma}_{\mu} } \Bigr\} f_{\sigma\nu}\nonumber\\
&& -\alpha_4 \Bigl\{ {\cal U}_3 {\cal K}^{\sigma}_{\mu} - {\cal U}_2 {\cal K}^{\rho}_{\mu}{\cal K}^{\sigma}_{\rho}
+ [{\cal K}]{\cal K}^{\rho}_{\mu}{\cal K}^{\delta}_{\rho}{\cal K}^{\sigma}_{\delta} \nonumber\\
&& - {\cal K}^{\rho}_{\mu}{\cal K}^{\delta}_{\rho} {\cal K}^{\gamma}_{\delta} {\cal K}^{\sigma}_{\gamma}\Bigr\}  f_{\sigma\nu} .
\end{eqnarray}
If we  introduce new variables:
\begin{eqnarray} \label{hatted-K}
&& \hat {\cal K}_{\mu\nu}= {\cal K}^{\sigma}_{\mu} f_{\sigma \nu}; ~\hat  {\cal K}_{\mu\nu}^2 = {\cal K}^{\rho}_{\mu}{\cal K}^{\sigma}_{\rho}f_{\sigma\nu}; ~\hat {\cal K}_{\mu\nu}^3= {\cal K}^{\rho}_{\mu}{\cal K}^{\delta}_{\rho}{\cal K}^{\sigma}_{\delta}  f_{\sigma\nu}; \nonumber\\
&& \hat {\cal K}^4_{\mu\nu}={\cal K}^{\rho}_{\mu}{\cal K}^{\delta}_{\rho} {\cal K}^{\gamma}_{\delta} {\cal K}^{\sigma}_{\gamma}  f_{\sigma\nu},
\end{eqnarray}
then the definition of $s_{\mu\nu}$ shown in Eq. (\ref{eq-s}) will be rewritten as \cite{SFH}
\begin{eqnarray}
s_{\mu\nu}(f) &=&  - \hat {\cal K}_{\mu\nu} + \left\{ [{\cal K}] +\alpha_3 {\cal U}_2 +\alpha_4 {\cal U}_3 \right\} f_{\mu\nu} \nonumber\\
&&+(\alpha_3+1)\left\{{ \hat {\cal K}_{\mu\nu}^2-[{\cal K}]\hat {\cal K}_{\mu\nu}}\right\}  \nonumber\\
&& -(\alpha_3+\alpha_4) \left\{{\hat {\cal K}_{\mu\nu}^3 -[{\cal K}]\hat {\cal K}_{\mu\nu}^2+{\cal U}_2 \hat {\cal K}_{\mu\nu} }\right\}    \nonumber\\
&& -\alpha_4 \left\{ {\cal U}_3 \hat{\cal K}_{\mu\nu} - {\cal U}_2 \hat {\cal K}_{\mu\nu}^2
+ [{\cal K}]\hat {\cal K}_{\mu\nu}^3 - \hat {\cal K}_{\mu\nu}^4 \right\}. \nonumber\\
\end{eqnarray}
 It is straightforward to see that if $f_{\mu\nu}=g_{\mu\nu}$ then we will obtain the following result:
\begin{eqnarray}
s_{\mu\nu} (f\to g) &=& - \left\{ X_{\mu\nu}(g)+ \alpha_4 \tilde Y_{\mu\nu}(g)+  \left({\cal U}_2 + \alpha_3 {\cal U}_3\right)g_{\mu\nu} \right\} \nonumber\\
& \equiv& t_{\mu\nu} ,
\end{eqnarray}
because $\hat {\cal K}_{\mu\nu}^n (f\to g) = {\cal K}_{\mu\nu}^n (g)$ with $n=1-4$. 

We would like to note that   the scalar curvature of metric $f_{\mu\nu}$ no longer shows up in the context of nonlinear massive gravity. Therefore, we only have a simple equation for  $f_{\mu\nu}$ in the dRGT theory:
\begin{equation} \label{equation-s}
s_{\mu\nu}^{\text{dRGT}}=0,
\end{equation}
provided that the unitary gauge of the St\"uckelberg scalar fields is chosen \cite{MSV,WFK}. As a result, in the dRGT theory the constraint equation, $s^{\text{dRGT}}_{\mu\nu}=0$, leads to  $t^{\text{dRGT}}_{\mu\nu}=0$, which will reduce the Einstein field equations to \cite{MSV,WFK}
\begin{equation}
M_g^2\left(R_{\mu\nu}(g)-\frac{1}{2}g_{\mu\nu}R(g)\right) - m^2  {\cal L}^{0,~\text{dRGT}}_M  g_{\mu\nu}=0 .
\end{equation}
 And the Bianchi identity indicates that the total massive graviton term acts as an effective cosmological constant because
\begin{equation}
\partial^\nu{\cal L}^{0,~\text{dRGT}}_M  =0.
\end{equation}
This is a general feature of the massive gravity. Once the forms of the physical and reference metrics are given, one can derive the following value of the effective cosmological constant in terms of the massive graviton terms as $\Lambda^0_M = -m^2  {\cal L}^{0,~\text{dRGT}}_M  $ as shown in {Refs.~}\cite{MSV,WFK}. 

However, in the bimetric gravity this result might not be valid for a large class of metrics due to the existence of the scalar curvature of metric $f_{\mu\nu}$. Note again that the reference metric in the bimetric gravity has been promoted as a dynamical one, in contrast to the dRGT theory. As a result, the appearance of  the scalar curvature $R(f)$ will lead to the differential equations (\ref{equation-f}) instead of the algebraic equations (\ref{equation-s}) for the reference metric $f_{\mu\nu}$. Hence, the constant-like behavior of the massive graviton terms in the context of the bimetric gravity theory could address more constraints. 
\section{Five-dimensional massive bigravity} \label{sec3}
Following the seminal paper of Hassan and Rosen on the four-dimensional massive bigravity \cite{SFH}, one can propose a consistent higher dimensional ($n>4$) massive bigravity as follows \footnote{After uploading the first version(s) of this paper to the arXiv website, the author has received a few comments claiming that the higher dimensional massive graviton terms of massive (bi)gravity have already been investigated in the published papers, e.g., Refs.~\cite{higher-bigravity,higher-bigravity-more}. Therefore, it is necessary to summarize briefly here what the author has done in Ref.~\cite{TQD} as well as in this paper in order to avoid some misunderstanding. In particular,  the author has explicitly shown  in Ref. ~\cite{TQD} that the massive graviton terms can be reconstructed from the characteristic equation of square matrix, which is a consequence of the Cayley-Hamilton theorem. As a result, this method turns out to be very effective in building up arbitrary dimensional graviton terms ${\cal L}_n$. This result indicates that there is indeed a close relation between the ghost-free property of the massive graviton terms and the Cayley-Hamilton theorem.  After constructing the higher dimensional ${\cal L}_n$'s ($n=5,6,7$), the author has compared them with that already derived in the published papers by other people  (see Secs. I and III of Ref. ~\cite{TQD} for more details). This clearly implies that the author has not been not among the first people \cite{higher-bigravity,higher-bigravity-more} investigating higher dimensional massive graviton terms. However, Ref.~\cite{TQD} and the present paper seem to be ones of the first papers studying explicitly nontrivial cosmological and black hole metrics such as the FLRW, Bianchi type I, and Schwarzschild-Tangherlini metrics for a specific five-dimensional massive (bi)gravity involving an additional graviton term ${\cal L}_5$ (or ${\cal U}_5$).}
\begin{eqnarray} \label{higher-action2}
S_{\text {nd}} &=& M_g^2 \int {d^n } x\sqrt { g} R(g)+ M_f^2 \int {d^n } x\sqrt {f} R(f) \nonumber\\
&&+2m^2 M_{\text{eff}}^2 \int {d^n } x\sqrt { g} \Bigl( {\cal U}_2 +\alpha_3 {\cal U}_3 +\alpha_4 {\cal U}_4 \nonumber\\
&&+\alpha_5 {\cal U}_5 +\alpha_6 {\cal U}_6 +\alpha_7 {\cal U}_7+...+\alpha_n{\cal U}_n\Bigr), \nonumber\\
\end{eqnarray}
where the first three higher dimensional graviton terms ($n=5,~6,~7$) are given by \cite{higher-bigravity,higher-bigravity-more,TQD}
\begin{eqnarray}
{\cal U}_5 &=&\frac{1}{120}\Bigl\{ [{\cal K}]^5 -10 [{\cal K}]^3 [{\cal K}^2] +20[{\cal K}]^2 [{\cal K}^3]   -20  [{\cal K}^2][{\cal K}^3] \nonumber\\
&&+15[{\cal K}]  [{\cal K}^2]^2 -30[{\cal K}] [{\cal K}^4]  +24 [{\cal K}^5] \Bigr\}, 
\end{eqnarray}
\begin{eqnarray}
{\cal U}_6&=& \frac{1}{720} \Bigl\{ [{\cal K}]^6 -15[{\cal K}]^4 [{\cal K}^2]+40[{\cal K}]^3 [{\cal K}^3] - 90 [{\cal K}]^2 [{\cal K}^4] \nonumber\\
&& +45 [{\cal K}]^2 [{\cal K}^2]^2  -15 [{\cal K}^2]^3 +40 [{\cal K}^3]^2 -120  [{\cal K}^3] [{\cal K}^2] [{\cal K}]   \nonumber\\
&&  +90[{\cal K}^4] [{\cal K}^2]   +144 [{\cal K}^5] [{\cal K}]-120 [{\cal K}^6] \Bigr\},
\end{eqnarray}
\begin{eqnarray}
\label{U7}
{\cal U}_7&=&\frac{1}{5040} \Bigl\{[{\cal K}]^7 -21 [{\cal K}]^5[{\cal K}^2]+70 [{\cal K}]^4[{\cal K}^3] -210[{\cal K}]^3 [{\cal K}^4] \nonumber\\
&&+105[{\cal K}]^3[{\cal K}^2]^2 -420[{\cal K}]^2[{\cal K}^2][{\cal K}^3]  +504[{\cal K}]^2[{\cal K}^5] \nonumber\\
&& -105[{\cal K}^2]^3[{\cal K}]+210[{\cal K}^2]^2 [{\cal K}^3]-504[{\cal K}^2][{\cal K}^5] \nonumber\\
&&+280[{\cal K}^3]^2[{\cal K}] -420[{\cal K}^3] [{\cal K}^4] +630[{\cal K}^4] [{\cal K}^2] [{\cal K}] \nonumber\\
&& -840[{\cal K}^6][{\cal K}]+720[{\cal K}^7] \Bigr\}.
\end{eqnarray}

For the method based on the Cayley-Hamilton theorem to construct higher dimensional interaction terms ${\cal U}_n$'s ($n>4$),  see {Ref.~}\cite{TQD} (see also the Appendix A in Ref.~\cite {review-bigravity}). Note that all graviton terms ${\cal U}_m$ with $m>n$ must vanish automatically in a given $n$-dimensional massive bigravity due to the requirement of absence of ghost   \cite{RGT,proof,higher-bigravity,higher-bigravity-more,TQD}. 
In this paper,   we will limit ourselves to the five-dimensional ($n=5$) massive bigravity described by the following action:
\begin{eqnarray} \label{action2}
S_{\text {5d}} &=& M_g^2 \int {d^5 } x\sqrt {g} R(g)+ M_f^2 \int {d^5 } x\sqrt {f} R(f) \nonumber\\
&&+2m^2 M_{\text{eff}}^2 \int {d^5 } x\sqrt { g} \Bigl( {\cal U}_2 +\alpha_3 {\cal U}_3 +\alpha_4 {\cal U}_4 +\alpha_5 {\cal U}_5 \Bigr). \nonumber\\
\end{eqnarray}
As a result, the corresponding five-dimensional Einstein field equations turn out to be \cite{higher-bigravity,higher-bigravity-more,TQD}
\begin{equation} \label{Einstein-5d}
M_g^2 \left({R_{\mu\nu}-\frac{1}{2}Rg_{\mu\nu}}\right)+m^2 M_{\text{eff}}^2 {\cal H}^{(5)}_{\mu\nu} (g)=0,
\end{equation}
where 
\begin{eqnarray} \label{def-of-H}
{\cal H}^{(5)}_{\mu\nu} (g)={ { X}_{\mu\nu}^{(5)}+ \sigma {Y}_{\mu\nu}^{(5)} +\alpha_5 {W}_{\mu\nu}} ,
\end{eqnarray}
\begin{eqnarray} \label{eqX-5d}
X_{\mu\nu}^{(5)}&=&  - \left(\alpha {\cal U}_2 +\beta {\cal U}_3 \right) g_{\mu\nu} + \tilde X_{\mu\nu}^{(5)},\\
\tilde X_{\mu\nu}^{(5)}&=& {\cal K}_{\mu\nu} -[{\cal K}]g_{\mu\nu} -\alpha\Bigl\{{{\cal K}_{\mu\nu}^2-[{\cal K}]{\cal K}_{\mu\nu} }\Bigr\} \nonumber\\
&&+\beta \Bigl\{{{\cal K}_{\mu\nu}^3-[{\cal K}]{\cal K}_{\mu\nu}^2+{\cal U}_2 {\cal K}_{\mu\nu} }\Bigr\}, 
\end{eqnarray}
\begin{eqnarray}
\label{eqY-5d}
 Y_{\mu\nu}^{(5)} &=& - {\cal U}_4 g_{\mu\nu} + \tilde Y_{\mu\nu}^{(5)},\\
\tilde Y_{\mu\nu}^{(5)} &=&{\cal U}_3 {\cal K}_{\mu\nu}  - {\cal U}_2  {\cal K}^2_{\mu\nu} +[{\cal K}]{\cal K}^3_{\mu\nu} -{\cal K}^4_{\mu\nu},
\end{eqnarray}
\begin{eqnarray}
W_{\mu\nu} &=& -{\cal U}_5g_{\mu\nu} + \tilde W_{\mu\nu}, \\
\tilde W_{\mu\nu} &=& {\cal U}_4  {\cal K}_{\mu\nu} - {\cal U}_3 {\cal K}^2_{\mu\nu}+{\cal U}_2 {\cal K}^3_{\mu\nu} - [{\cal K}]{\cal K}^4_{\mu\nu} +{\cal K}^5_{\mu\nu}, \nonumber\\
\end{eqnarray}
with $\alpha = \alpha_3+1$, $\beta =\alpha_3+\alpha_4$, and $\sigma =\alpha_4+\alpha_5$ are additional parameters defined for convenience.

It is noted that in the four-dimensional spacetime, where $W_{\mu\nu}$ no longer exists, the tensor $Y_{\mu\nu}$ has been shown to be zero as a consequence of the Cayley-Hamilton theorem. And in five-dimensional spacetime, we also have the same result for the tensor $W_{\mu\nu}$, i.e., $W_{\mu\nu}=0$ in general as a consequence of the Cayley-Hamilton theorem. Note again that $Y_{\mu\nu}\neq 0$ in the higher dimensional  ($n > 4$) spacetime. Of course, $W_{\mu\nu}\neq 0$ in higher-than-five dimensional spacetimes.

For the reference metric, $f_{\mu\nu}$, its Einstein field equations can be defined to be \cite{higher-bigravity,higher-bigravity-more}
\begin{equation} \label{5d-equation-f}
\sqrt {f} M_f^2\left(R_{\mu\nu}(f)-\frac{1}{2}f_{\mu\nu}R(f)\right)+ \sqrt{g} m^2 M_{\text{eff}}^2 s_{\mu\nu}^{(5)}(f) =0,
\end{equation}
where the tensor $s_{\mu\nu}^{(5)}(f) $ is given by
\begin{eqnarray} \label{5d-def-of-s}
s_{\mu\nu}^{(5)}(f) &\equiv& - {\hat{\cal K}}_{\mu\nu} +\Bigl\{ [{\cal K}]+ \alpha_3 {\cal U}_2 +\alpha_4 {\cal U}_3 +\alpha_5{\cal U}_4\Bigr\} f_{\mu\nu} \nonumber\\
&& + \alpha \left\{{{\hat{\cal K}}_{\mu\nu}^2-[{\cal K}]{\hat{\cal K}}_{\mu\nu} }\right\} \nonumber\\
&& -\beta \left\{{{\hat{\cal K}}_{\mu\nu}^3-[{\cal K}]{\hat{\cal K}}_{\mu\nu}^2+{\cal U}_2 {\hat{\cal K}}_{\mu\nu} }\right\} \nonumber\\
&& - \sigma \left\{{\cal U}_3 {\hat{\cal K}}_{\mu\nu}  - {\cal U}_2  {\hat{\cal K}}^2_{\mu\nu} +[{\cal K}]{\hat{\cal K}}^3_{\mu\nu} -{\hat{\cal K}}^4_{\mu\nu} \right\} \nonumber\\
&& -\alpha_5 \left\{ {\cal U}_4  {\hat{\cal K}}_{\mu\nu} - {\cal U}_3 {\hat{\cal K}}^2_{\mu\nu}+{\cal U}_2 {\hat{\cal K}}^3_{\mu\nu} - [{\cal K}]{\hat{\cal K}}^4_{\mu\nu} \right.\nonumber\\
&&\left. +{\hat{\cal K}}^5_{\mu\nu} \right\}.
\end{eqnarray}
Here ${\hat{\cal K}}^5_{\mu\nu} ={\cal K}^{\rho}_{\mu}{\cal K}^{\delta}_{\rho} {\cal K}^{\gamma}_{\delta} {\cal K}^{\sigma}_{\gamma}  {\cal K}^{\alpha}_{\sigma} f_{\alpha\nu}$ and  ${\hat{\cal K}}^n_{\mu\nu}$'s ($n=1-4$) have been defined in Eq. (\ref{hatted-K}).

Along with the physical and reference metric equations defined above, there are two Bianchi identities for the physical and reference metrics:
\begin{eqnarray}\label{Bianchi-identity-for-g}
D^\mu_g G_{\mu\nu}(g)=0,\\
\label{Bianchi-identity-for-f}
D^\mu_f G_{\mu\nu}(f)=0,
\end{eqnarray}
respectively, which lead to the following constraint equations:
\begin{eqnarray} \label{constraint1}
D^\mu_g {\cal H}^{(5)}_{\mu\nu}(g)=0,\\
 \label{constraint2}
D^\mu_f \left[ \frac{\sqrt{g}}{\sqrt{f}}s_{\mu\nu}^{(5)}(f) \right]=0.
\end{eqnarray}
Here, notations $D^\mu_g$ and $D^\mu_f$ stand for the covariant derivatives in the $g$- and $f$-sectors, respectively.

By following the same technique shown above, one can extend the bigravity to higher-than-five-dimensional scenarios. For example, one can propose to work with six- and seven-dimensional bigravity models corresponding to six and seven-dimensional graviton terms, which have been constructed in the context of higher non-linear massive gravity~\cite{higher-bigravity,higher-bigravity-more,TQD}. Note also that  we have shown in the recent work~\cite{TQD} that the five-dimensional dRGT theory admits the constant-like property of massive graviton terms for a number of physical and compatible reference metrics  as the FLRW, Bianchi type I, and Schwarzschild-Tangherlini (anti-) de Sitter [(A)dS] black holes metrics. This result is basically based on  the non-dynamical  feature of the reference metric $f_{ab}$. When the reference metric becomes dynamical in the framework of bigravity, however, we might need additional constraints in order to make the massive graviton terms  constant. Hence, the Einstein field  equations of higher dimensional bigravity will be more complicated to solve if such constraints are not introduced. In the next sections, we will try to seek the above mentioned metrics for the five-dimensional bigravity under an assumption that the physical metrics are compatible with the physical ones. It appears that the compatibility assumption can be reduced to a special case, which leads to constant-like behavior of ${\cal U}_M$,  such that the reference metrics are proportional  to the physical ones, i.e., $f_{\mu\nu}={\cal C}^2 g_{\mu\nu}$ with a constant ${\cal C}$, similar to the earlier works, e.g., \cite{SFH,Hassan:2012wr,review-bigravity,higher-bigravity,higher-bigravity-more}. 
\section{Five-dimensional FLRW metrics} \label{sec4}
\subsection{Field equations}
In five-dimensional spacetimes, we will consider the  FLRW physical and reference metrics  given by 
\begin{eqnarray} \label{frw-1}
ds_{\text {5d}}^2 (g_{\mu\nu}) &=& -N_1^2(t)dt^2 + a_1^2(t) \left(d\vec{x}^2+du^2 \right), \\
 \label{frw-2}
ds_{\text {5d}}^2 (f_{\mu\nu}) &=& -N_2^2(t)dt^2 + a_2^2(t) \left(d\vec{x}^2 +du^2 \right),
\end{eqnarray}
where $a_i$'s ($i=1-2$) are the scale factors  and $u$ is the fifth dimension \cite{5d-FRW}.  In addition, $N_1$ and $N_2$ are the lapse functions, which are introduced to result the Friedmann equations from their Euler-Lagrange equations \cite{RGT,SFH}. Furthermore, $N_1$ can be set to be one after obtaining the following Friedmann equation. However, we should not do the same thing for $N_2$, i.e., $N_2$ should be regarded as a free field variable.  For convenience, we will define additional variables \cite{TQD}, which will be used for further calculations, as follows
\begin{eqnarray}
[{\cal K}]^n &=& \left( \gamma + 4\Sigma\right)^n, ~  [{\cal K}^n] = \gamma^n +4\Sigma^n, \nonumber\\
{\hat{\cal K}}^n_{00}&=&\gamma^n f_{00},~ {\hat{\cal K}}^n_{ii}=\Sigma^n f_{ii},~ \frac{\sqrt{f}}{\sqrt{g}}= (1-\gamma)(1- \Sigma)^4,\nonumber\\
\gamma (t)&=&1- \frac{N_2(t)}{N_1(t)}, ~ \Sigma(t) =1- \frac{a_2(t)}{a_1(t)}.
\end{eqnarray}

Thanks to these definitions, the massive graviton terms ${\cal U}_i$'s can be reduced to
\begin{eqnarray}
{\cal U}_2&=&2 \Sigma  \left(2 \gamma +3 \Sigma \right), \\
{\cal U}_3&=&2 \Sigma^2 \left(3 \gamma+2 \Sigma \right),\\
{\cal U}_4&=&\Sigma^3 \left(4 \gamma+\Sigma \right), \\
{\cal U}_5&=&\gamma \Sigma^4,
\end{eqnarray}
which will be used to evaluate the total graviton term ${\cal U}_M$ to be
\begin{eqnarray} \label{UM}
{\cal U}_M &\equiv& {\cal U}_2+\alpha_3 {\cal U}_3+\alpha_4 {\cal U}_4 +\alpha_5 {\cal U}_5 \nonumber\\
&=&\Sigma \left[ \left(\alpha _5\gamma+\alpha _4 \right)\Sigma ^3 +4 \left(\alpha _4 \gamma+\alpha _3 \right) \Sigma ^2 \right. \nonumber\\
 &&\left.+6  \left(\alpha _3 \gamma +1\right)\Sigma+4\gamma \right].
\end{eqnarray}
Hence,  the non-vanishing components of the tensor ${\cal H}^{(5)}_{\mu\nu}$ defined in Eq. (\ref{def-of-H}) are given by
\begin{eqnarray}
{\cal H}^{(5)}_{00} &=& -\Sigma \left(\sigma \Sigma^3 + 4\beta \Sigma^2 +6\alpha \Sigma +4 \right) g_{00},\\
{\cal H}^{(5)}_{ii} &=&- \left[ \gamma \left(\sigma \Sigma^3+3\beta \Sigma^2+3\alpha \Sigma +1\right) \right. \nonumber\\
&& \left. +\Sigma \left(\beta\Sigma^2+3\alpha \Sigma +3\right) \right] g_{ii}.
\end{eqnarray}
Additionally, the non-vanishing components of the tensor $s_{\mu\nu}^{(5)}$ shown in Eq. (\ref{5d-def-of-s}) turn out to be
\begin{align}
s_{00}^{(5)}=&~\left(1-\gamma \right) \Sigma \left(\alpha _5 \Sigma ^3+4 \alpha _4 \Sigma ^2+6 \alpha _3 \Sigma +4\right) f_{00},\\
s_{ii}^{(5)}=&~\left(1-\Sigma \right) \left[ \left(\alpha _5\gamma+\alpha _4 \right)\Sigma ^3 +3 \left(\alpha _4 \gamma+\alpha _3 \right) \Sigma ^2 \right.\nonumber\\
& \left. +3  \left(\alpha _3 \gamma +1\right)\Sigma+\gamma \right] f_{ii}.
\end{align}

Armed with these results,  the Einstein field equations (\ref{Einstein-5d}) now become as
\begin{align} \label{physical-eq-1}
&6\tilde M_g^2 H_1^2 +\Sigma\left(\sigma \Sigma^3 + 4\beta \Sigma^2 +6\alpha \Sigma +4 \right)=0,\\
\label{physical-eq-2}
& 3 \tilde M_g^2 \left(\dot H_1 +2 H_1^2 \right) +\gamma \left(\sigma \Sigma^3+3\beta \Sigma^2+3\alpha \Sigma +1\right)  \nonumber\\
&+\Sigma \left(\beta\Sigma^2+3\alpha \Sigma +3\right)=0, 
\end{align}
where  $H_1 =\dot a_1/a_1$ is the Hubble constant for the physical metric $g_{\mu\nu}$. In addition, we have set an additional variable as $\tilde M_g^2 \equiv M_g^2/(m^2 M_{\text{eff}}^2)$ and $N_1=1$ has been chosen for convenience.
On the other hand,  the reference metric equations (\ref{5d-equation-f}) turn out to be
\begin{align} \label{reference-eq-1}
&6\tilde M_f^2 \left(1-\Sigma \right)^4 H_2^2 \nonumber\\
& -\Sigma \left(1-\gamma\right)^2 \left(\alpha _5 \Sigma ^3+4 \alpha _4 \Sigma ^2+6 \alpha _3 \Sigma +4\right) =0 ,\\
\label{reference-eq-2}
&3 \tilde M_f^2 \left(1-\Sigma\right)^3 \left(\dot H_2 +2 H_2^2 +\frac{\dot\gamma}{1-\gamma}H_2\right) \nonumber\\
& - \left(1-\gamma \right)  \left[ \left(\alpha _5\gamma+\alpha _4 \right)\Sigma ^3 +3 \left(\alpha _4 \gamma+\alpha _3 \right) \Sigma ^2 \right. \nonumber\\
&\left. +3  \left(\alpha _3 \gamma +1\right)\Sigma+\gamma \right] =0,
\end{align}
where $H_2 =\dot a_2/a_2$ the Hubble constant for the reference metric $f_{\mu\nu}$, $\tilde M_f^2 \equiv M_f^2/(m^2 M_{\text{eff}}^2)$,  and $\gamma =1-N_2$ due to the setting $N_1=1$.

Along with these field equations, the constraint equations (\ref{constraint1}) and (\ref{constraint2}) turn out to be
\begin{align}
g^{00}\partial_0 {\cal H}^{(5)}_{00} -4 g^{11} \left[\Gamma^{0}_{11}(g){\cal H}_{00}^{(5)}+\Gamma^{1}_{10}(g){\cal H}_{11}^{(5)}\right]=0
\end{align}
and
\begin{align}
& f^{00} \left\{ \partial_0 \left[ \frac{\sqrt{g}}{\sqrt{f}}s_{00}^{(5)}\right] -2 \frac{\sqrt{g}}{\sqrt{f}}\Gamma^{0}_{00}(f) s_{00}^{(5)} \right\} \nonumber\\
& -4 \frac{\sqrt{g}}{\sqrt{f}}f^{11} \left[ \Gamma^{0}_{11}(f)s_{00}^{(5)} +\Gamma^{1}_{10}(f) s_{11}^{(5)}\right]=0, 
\end{align}
respectively. Furthermore, these equations can be reduced to
\begin{align} \label{constraint3}
&\partial_0 \left[\Sigma \left(\sigma \Sigma^3 + 4\beta \Sigma^2 +6\alpha \Sigma +4 \right)\right] \nonumber\\
&=-4 H_1 \left(\Sigma-\gamma\right) \left(\sigma \Sigma^3+3\beta \Sigma^2+3\alpha \Sigma +1\right) 
\end{align}
and
\begin{align}\label{constraint4}
&  \left(4H_1 -4H_2 +\frac{3\dot\gamma}{1-\gamma} +\partial_0 \right)\nonumber\\
& \times\left(1-\gamma \right) \Sigma \left(\alpha _5 \Sigma ^3+4 \alpha _4 \Sigma ^2+6 \alpha _3 \Sigma +4\right) \nonumber\\
&=- 4H_2 \left(\Sigma-\gamma\right) \left(\sigma \Sigma^3+3\beta \Sigma^2+3\alpha \Sigma +1\right),
\end{align}
respectively. 

It appears that the set of constraint equations (\ref{constraint3}) and (\ref{constraint4}) is quite complicated to solve analytically generally. However,  simple solutions can be figured out from this set equations if two right-hand sides of Eqs. (\ref{constraint3}) and (\ref{constraint4}) vanish altogether.  As a result, this assumption can be achieved with one of two possible cases:
\begin{align} \label{FRW-case1}
&\text{(i)}~\gamma=\Sigma,\\
 \label{FRW-case2}
&\text{(ii)}~\sigma \Sigma^3+3\beta \Sigma^2+3\alpha \Sigma +1 = 0,
\end{align}
given that $H_i  \neq 0$ ($i=1-2$).
Consequently, this assumption also implies that 
\begin{equation}
\Sigma \left(\sigma \Sigma^3 + 4\beta \Sigma^2 +6\alpha \Sigma +4 \right) = \text {constant}.
\end{equation}
This constancy property indicates that $\Sigma$ should also be constant consistently since $\alpha$, $\beta$, and $\sigma$ all are constant coefficients. In addition, the constancy of $\Sigma$, i.e., $\dot \Sigma =0$, indicates that
\begin{equation}
 H_1 =H_2. 
\end{equation}
On the other hand,  Eq. (\ref{constraint4}) leads to
\begin{equation}
\dot\gamma= 0.
\end{equation}

 Therefore, this result also implies that $\gamma$ acts as a constant, similar to $\Sigma$.
More interestingly, the total massive graviton term will also act as an effective cosmological constant $\Lambda_M$ due to the constancy feature of $\gamma$ and $\Sigma$: 
\begin{equation} \label{LM}
\Lambda_M =- m^2 M_{\text{eff}}^2 {\cal U}_M,
\end{equation}
where the definition of the total massive graviton term ${\cal U}_M$ has been defined in Eq. (\ref{UM}). It is apparent that once the values of $\gamma$ and $\Sigma$ are solved, the corresponding value of $\Lambda_M$ will be determined in terms of that of ${\cal U}_M$ as shown in Eq. (\ref{LM}). In the following subsections, therefore, we will consider separately  two possible cases shown in Eqs. (\ref{FRW-case1}) and (\ref{FRW-case2}) in order to solve the equations of physical and reference metrics as defined in Eqs. (\ref{physical-eq-1}), (\ref{physical-eq-2}), (\ref{reference-eq-1}), and (\ref{reference-eq-2}). Additionally, we will show how to compute the value of $\gamma$ along with that of $\Sigma$ by  deriving their corresponding equations. 
\subsection{Analytical solutions}
\subsubsection{Case 1: $\gamma =\Sigma$}
As shown above, the variables $\Sigma$ and $\gamma$ should be constant altogether. Hence, we will assume that
\begin{equation}
\gamma= \Sigma = \hat C =\text{constant}.
\end{equation} 
It appears that this assumption is equivalent to the choice, which has been taken in many earlier works in the context of bigravity theory \cite{Hassan:2012wr,review-bigravity,higher-bigravity,higher-bigravity-more}, that the reference metric is proportional to the physical metric:
\begin{equation}
f_{\mu\nu}= (1-{\hat C})^2  g_{\mu\nu}.
\end{equation}

Our goal is to define the value of the constant $\hat C$, which has been assumed to be identical to that of $\gamma$ and $\Sigma$, by considering the reference and physical field equations. As a result, Eqs. (\ref{physical-eq-1}) and (\ref{physical-eq-2}) imply that
\begin{eqnarray}
\label{physical-eq-3}
\dot H_1 &=&0, \\ 
\label{physical-eq-4}
6  H_1^2 &=& {\hat\Lambda}^g_0 ,
\end{eqnarray}
where ${\hat\Lambda}^g_0$ is an effective cosmological constant for the $g$-sector associated with the physical metric $g_{\mu\nu}$ defined as follows
\begin{align}
\frac{1}{\tilde M_g^2} {\cal H}^{(5)}_{\mu\nu} &= {\hat\Lambda}^g_0 g_{\mu\nu},\\
\label{L0}
{\hat\Lambda}^g_0 &= -\frac{\hat C}{\tilde M_g^2}\left(\sigma \hat C^3 + 4\beta \hat C^2 +6\alpha \hat C +4 \right).
\end{align}
On the other hand, Eqs. (\ref{reference-eq-1}) and (\ref{reference-eq-2}) lead to
\begin{eqnarray}
\label{reference-eq-3}
\dot H_2 &=&0, \\
\label{reference-eq-4}
6 H_2^2 &=& (1-\hat C)^2 {\hat \Lambda}^f_0,
\end{eqnarray}
where ${\hat\Lambda}^f_0$ is an effective cosmological constant for the $f$-sector associated with the reference metric $f_{\mu\nu}$ given by
\begin{align}
\frac{\sqrt{g}}{{\sqrt{f}}\tilde M_f^2} s^{(5)}_{\mu\nu} &= {\hat\Lambda}^f_0 f_{\mu\nu},\\
\label{LF}
{\hat\Lambda}^f_0 &= \frac{ \hat C}{ \tilde M_f^2 (1-\hat C)^4} \left(\alpha _5 \hat C ^3+4 \alpha _4 \hat C ^2+6 \alpha _3 \hat C +4\right).
\end{align}

It is noted that we have derived the relation, $H_1 =H_2$, from the constancy property of $\Sigma$. Hence, we obtain the following relation from Eqs. (\ref{physical-eq-4}) and (\ref{reference-eq-4}) such as  ${\hat\Lambda}^g_0=(1-\hat C)^2{\hat\Lambda}^f_0$, which can be expanded to give a degree 5 polynomial equation of $ \hat C$:
\begin{eqnarray} \label{eq-of-hat-C}
 &&\sigma \hat C^5 -2 \left(\sigma-2\beta \right) \hat C^4 + \left(\sigma-8\beta + 6\alpha + \alpha_5 \tilde M^2 \right)\hat C^3 \nonumber\\
&& + 4\left(\beta -3\alpha +\alpha_4 \tilde M^2+1\right)\hat C^2 \nonumber\\
 &&+2\left(3\alpha+3\alpha_3\tilde M^2-4\right) \hat C +4 \left( \tilde M^2+1 \right) =0,
\end{eqnarray}
with $\tilde M^2 \equiv \tilde M_g^2 / \tilde M_f^2$ as a dimensionless parameter.  Mathematically, this polynomial equation admits five real or complex solutions of $\hat C$. Physically, however, $\hat C$ should be real definite for expanding physical and reference metrics. Solving Eq. (\ref{eq-of-hat-C}) will yield the corresponding real values of $\hat C$. In the dRGT limit, where $s^{(5)}_{\mu\nu}=0$ due to the non-dynamical property of the reference metric $f_{\mu\nu}$, the corresponding equation of $\hat C$ turns out to be
\begin{equation}
\alpha_5 \hat C^3 +4\alpha_4 \hat C^2 +6\alpha_3 \hat C +4=0,
\end{equation}
which is identical to that investigated for the FLRW metrics in \cite{TQD}. 
 
As a result, integrating out both Eqs. (\ref{physical-eq-3}) and (\ref{reference-eq-3}) implies that
\begin{equation} \label{a1}
a_ 1 (t) = \exp \left[\sqrt{\frac{\hat\Lambda_0^g}{6}} t \right];~
a_2 (t) = (1- \hat C) \exp \left[\sqrt{\frac{\hat\Lambda_0^g}{6}} t \right].
\end{equation}
It is apparent that these solutions will be exactly the de-Sitter expanding solution if $\hat\Lambda_0^g>0$.
In this case, the effective cosmological constant $\Lambda_M$, which is associated with the total massive graviton term ${\cal U}_M$  in Eq. (\ref{UM}) as shown in Eq. (\ref{LM}), can be evaluated to be
\begin{align} \label{lambdaM_expression}
\Lambda_M &= -m^2 M_{\text{eff}}^2 \hat C \left[\left(\sigma \hat C^3 + 4\beta \hat C^2 +6\alpha \hat C +4 \right) \right.\nonumber\\
& \left.+\left(\hat C-1\right)\left(\alpha _5 \hat C ^3+4 \alpha _4 \hat C ^2+6 \alpha _3 \hat C +4\right) \right] \nonumber\\
&= \hat\Lambda_0^g \left[M_g^2 +M_f^2 (1-\hat C)^3\right].
\end{align}

The expression of $\Lambda_M$ shown in Eq. (\ref{lambdaM_expression}) indicates that it is combined from two effective cosmological constants $\hat\Lambda_0^g$ and $\hat\Lambda_0^f$ defined in the $g$- and $f$- sectors, respectively.  In particular, if $M_g^2 >M_f^2(\hat C-1)^3 $ then $\Lambda_M >0$ and vice versa, provided that $\hat \Lambda_0^g >0$. Hence, it turns out that $\hat\Lambda_0^g \neq \Lambda_M/M_g^2$ in the context of the massive bigravity theory. This is a different point of the massive bigravity  compared with the massive gravity. It has been shown in the dRGT theory that $\hat\Lambda_0^g$ should be identical to $\Lambda_M/M_g^2$ if the physical metric is taken to be compatible with the reference metric \cite{TQD,WFK}.  In the bigravity theory, however, there exists the Ricci scalar $R(f)$ due to the   assumption of dynamical reference metric $f_{\mu\nu}$, which leads to the differential field equations of reference scale factors $\alpha_2$, $\sigma_2$, and $\beta_2$. Hence, in the expression of $\Lambda_M$ as shown in Eq. (\ref{lambdaM_expression}), there are terms associated with $f_{\mu\nu}$, which can be set equal to zero if $f_{\mu\nu}$ is non-dynamical, or equivalently $R(f)=0$. Indeed, in the case of absence of $R(f)$, the field equations for the reference metric $f_{\mu\nu}$ will be $s^{(5)}_{\mu\nu}(f)=0$ rather than Eq. (\ref{5d-equation-f}). The case of non-dynamical reference $f_{\mu\nu}$ is nothing but the massive gravity theory, which has been investigated in {Ref.~}\cite{TQD} for a number of  metrics including the five-dimensional FLRW one. 

Now, we would like to see whether both effective cosmological constants, ${\hat\Lambda}^f_0$ and ${\hat\Lambda}^g_0$, and of course $\Lambda_M$, vanish. It turns out that if $\alpha_i$'s ($i=3-5$) satisfy both the following equations:
\begin{eqnarray}
\sigma \hat C^3 + 4\beta \hat C^2 +6\alpha \hat C +4&=&0,\\
\alpha _5 \hat C ^3+4 \alpha _4 \hat C ^2+6 \alpha _3 \hat C +4 &=&0,
\end{eqnarray}
then ${\hat\Lambda}^f_0={\hat\Lambda}^g_0=0$ as expected. As a result, these two constraint equations can be rewritten as follows
\begin{eqnarray}\label{hat-C-eq1-lambda-zero}
\alpha _4 \hat C ^2+3 \alpha _3 \hat C +3 &=&0,\\
\label{hat-C-eq2-lambda-zero}
\alpha_5 \hat C^3 +3\alpha_4 \hat C^2 +3\alpha_3 \hat C+1&=&0.
\end{eqnarray}
As a result, solving the first equation (\ref{hat-C-eq1-lambda-zero}) gives us non-trivial solutions of $\hat C$:
\begin{equation}
{\hat C}=\frac{-3 \alpha _3 \pm \sqrt{3\left( 3\alpha _3^2-4 \alpha _4\right)}}{2 \alpha _4},
\end{equation}
requiring that $\alpha_3^2 >(4/3) \alpha_4$. Hence, the corresponding value of $\alpha_5$ can be defined from the second equation (\ref{hat-C-eq2-lambda-zero})  to be
\begin{align} \label{special-a5}
\alpha_5 =&~- \frac{3\alpha_4 {\hat C}^2 +3 \alpha_3 {\hat C} +1}{{\hat C}^3} \nonumber\\
=&~ \frac{8\alpha _4^2 \left[\left(9 \alpha _3^2-8 \alpha _4\right)\mp3 \alpha _3 \sqrt{9 \alpha _3^2-12 \alpha _4}  \right] }{\left(3 \alpha _3\mp\sqrt{9 \alpha _3^2-12 \alpha _4}\right)^3}. 
\end{align}

In short, once the real value of $\hat C$, or equivalently that of $\Sigma$ and $\gamma$, is solved from the following polynomial equation (\ref{eq-of-hat-C}), the value of $\hat\Lambda_0^g$ will be evaluated according to Eq. (\ref{L0}). Consequently, the scale factors $a_1$ and $a_2$ of the FLRW physical and reference metrics will be determined as shown in Eq. (\ref{a1}).
\subsubsection{Case 2: $\sigma \Sigma^3+3\beta \Sigma^2+3\alpha \Sigma +1=0$}
As a result,   the physical metric equations (\ref{physical-eq-1}) and (\ref{physical-eq-2}) both imply, under this case, that
\begin{eqnarray}
\dot H_1 &=&0, \\
6 \tilde M_g^2 H_1^2 &=& -\Sigma\left(\beta \Sigma^2 +3\alpha \Sigma +3 \right).
\end{eqnarray}
On the other hand, the reference metric equations (\ref{reference-eq-1}) and (\ref{reference-eq-2}) can be solved to admit a trivial solution:
\begin{align}
\dot H_2 =&~0, \\
6 \tilde M_f^2 H_2^2 =&~\frac{\Sigma \left(1-\gamma\right)^2}{\left(1-\Sigma\right)^4} \left(\alpha _5 \Sigma ^3+4 \alpha _4 \Sigma ^2+6 \alpha _3 \Sigma +4\right).
\end{align}
As a result,  we can derive the following equation for $\gamma$ in terms of $\Sigma$ by using the fact that $H_1^2= H_2^2$.
For convenience, we rewrite ${\cal H}_{00}^{(5)}$ in this case as follows
\begin{eqnarray}
{\cal H}_{00}^{(5)} &=& \Lambda_g^{(5)} g_{00},\\
\Lambda_g^{(5)} &=&-\Sigma\left(\beta \Sigma^2 +3\alpha \Sigma +3 \right),
\end{eqnarray}
where $\Lambda_g^{(5)}$ acts as an effective cosmological constant. On the other hand,  we obtain the following result in the four-dimensional bigravity that
\begin{eqnarray}
{\cal H}_{00} &=& \Lambda_g^{(4)} g_{00},\\
 \Lambda_g^{(4)} &=&-\Sigma \left(\beta\Sigma^2 +3\alpha \Sigma +3 \right),
\end{eqnarray}
with an effective cosmological constant,  $\Lambda_g^{(4)}$. It turns out that if the equation, $\sigma \Sigma^3+3\beta \Sigma^2+3\alpha \Sigma +1=0$, holds then the five-dimensional effective cosmological constant  $\Lambda_g^{(5)}$ will recover the four-dimensional one $\Lambda_g^{(4)}$. Moreover, we will show later that this result is also valid for the case of proportional metrics.  Note that the case of proportional metrics has been studied in a number of previous papers, e.g., see \cite{SFH,Hassan:2012wr,review-bigravity,higher-bigravity,higher-bigravity-more,blackholes-review}. Hence, in the rest of paper, we will focus on the cases, where the reference metrics are taken to be proportional to the physical ones.
\section{Five-dimensional Bianchi type I metrics} \label{sec5}
\subsection{Field equations}
Now, we would like to go beyond the isotropic spacetime scenario, i.e., considering anisotropic spacetimes to see whether the five-dimensional bigravity model admits them as its cosmological solutions.  In a five-dimensional spacetime scenario, the Bianchi type I physical and reference metrics are taken to be \cite{TQD,5d-bianchi}
\begin{eqnarray}
\label{eq7}
 ds_{\text {5d}}^2 (g_{\mu\nu}) &= & -N_1^2(t)dt^2+\exp\left[{2\alpha_1(t)-4\sigma_1(t)}\right]dx^2\nonumber \\  &&+\exp\left[{2\alpha_1(t)+2\sigma_1(t)}\right]\left({dy^2+dz^2}\right) \nonumber\\
&&+\exp\left[{2\beta_1(t)}\right] du^2 , \\
 \label{eq8}
 ds_{\text {5d}}^2 (f_{\mu\nu})  &= &  -N_2^2(t)dt^2+\exp\left[{2\alpha_2(t)-4\sigma_2(t)}\right]dx^2\nonumber \\  &&+\exp\left[{2\alpha_2(t)+2\sigma_2(t)}\right]\left({dy^2+dz^2}\right) \nonumber\\
&& +\exp\left[{2\beta_2(t)}\right] du^2 , 
\end{eqnarray}
where $\beta_i$'s ($i=1-2$) are additional scale factors associated with the fifth dimension $u$ \cite{5d-bianchi}. Now, we would like to see if the five-dimensional bigravity admits these Bianchi type I metrics as its anisotropic cosmological solutions. For convenience, we will define some  useful definitions \cite{TQD,WFK}:
\begin{align}
[{\cal K}]^n =&~ \left(\gamma+A+2B+C\right)^n; \nonumber\\
 [{\cal K}^n] =& ~\gamma^n +A^n +2B^n+C^n; \nonumber\\
\gamma =&~1-\frac{N_2}{N_1};~A=1-\epsilon \eta^{-2};~B=1-\epsilon \eta; \nonumber\\
 C =&~1-\exp\left[\beta_2 -\beta_1\right];~\epsilon = \exp\left[\alpha_2 -\alpha_1\right]; \nonumber\\
\eta=&~\exp\left[\sigma_2 -\sigma_1\right]; \hat {\cal K}^n_{00} = \gamma^n f_{00};~ \hat {\cal K}^n_{11} = A^n f_{11}; \nonumber\\
\hat {\cal K}^n_{22}=&~\hat {\cal K}^n_{33} = B^n f_{33}; ~ \hat {\cal K}^n_{44} = C^n f_{44},
\end{align}
which will help us to reduce some complicated expressions of field and constraint equations to simple ones. In particular,  the massive graviton terms ${\cal U}_i$'s can be explicitly expanded  in terms of the above notations to be
\begin{align}\label{eq9}
{\cal U}_2 &= B \left(2A+B\right)+C\left(A+2B\right) +\gamma \left(A+2B+C\right),\\
\label{eq10}
{\cal U}_3 &=AB^2+B \left(2A+B\right)\left(\gamma+C\right) +\gamma C \left(A+2B\right),\\
\label{eq11}
{\cal U}_4 &= B \left[AB \left(\gamma+C\right) +\gamma C \left(2A+B\right)\right],\\
\label{eq12}
{\cal U}_5 &= \gamma A B^2 C.
\end{align}
In addition, the total massive graviton term ${\cal U}_M \equiv  {\cal U}_2 +\alpha_3 {\cal U}_3+\alpha_4 {\cal U}_4+\alpha_5 {\cal U}_5$ turns out to be
\begin{align}\label{eq13}
{\cal U}_M &= AB^2C\left(\alpha_5 \gamma +\alpha_4\right) \nonumber\\
&+B \left(\alpha_4 \gamma +\alpha_3 \right)\left[AB +C \left(2A+B\right)\right] \nonumber\\
&+\left(\alpha_3 \gamma+1\right) \left[B\left(2A+B\right)+C\left(A+2B\right)\right] \nonumber\\
&+\gamma \left(A+2B+C\right).
\end{align}
It is straightforward to see that in the isotropic limit, i.e., the FLRW limit corresponding to the case $A=B=C$, the above massive graviton terms all reduce to that defined for the FLRW metric in the previous section.

Given the above results, we arrive at the non-vanishing components of the tensor ${\cal H}_{\mu\nu}^{(5)}$ appearing in the physical metric field equation as displayed in Eq. (\ref{def-of-H}):
\begin{align}
{\cal H}_{00}^{(5)} & =  -\left\{ \sigma AB^2C +\beta B \left(AB+2AC+BC\right)  \right. \nonumber\\
&\left.+\alpha \left[B\left(2A+B\right)+C\left(A+2B\right)\right] \right. \nonumber\\
&\left. +A+2B+C \right\} g_{00}, \\
{\cal H}_{11}^{(5)} &=  - \left\{ \gamma \left[\sigma B^2 C +\beta B \left(B+2C\right) +\alpha \left(2B+C\right) +1\right]  \right. \nonumber\\
& \left. +\left[\beta B^2 C +\alpha B \left(B+2C\right) +2B+C \right] \right\}g_{11},\\
{\cal H}_{22}^{(5)} & =  - \left\{ \gamma \left[\sigma ABC + \beta \left(AB+AC+BC\right) \right.\right.\nonumber\\
&\left.\left. +\alpha \left(A+B+C\right)+1 \right] \right.\nonumber\\
&\left. +\left[\beta ABC +\alpha \left(AB+AC+BC\right)\right.\right.\nonumber\\
&\left.\left.  +A+B+C\right]\right\} g_{22},\\
{\cal H}_{33}^{(5)}  & =~ {\cal H}_{22}^{(5)},\\
{\cal H}_{44}^{(5)}  & = - \left\{\gamma\left[\sigma AB^2 +\beta B \left(2A+B\right) +\alpha \left(A+2B\right)+1\right] \right. \nonumber\\
& \left. +\left[\beta AB^2 +\alpha B \left(2A+B\right)+A+2B\right]\right\} g_{44}.
\end{align}
On the other hand, the non-vanishing components of the tensor $s_{\mu\nu}^{(5)}$ appearing in the reference metric field equation can be shown to be
\begin{align}
s_{00}^{(5)} &= ~ \left(1-\gamma\right) \left\{\alpha_5 AB^2 C +\alpha_4 B \left[AB+2AC +BC\right] \right. \nonumber\\
&\left.+\alpha_3 \left[ B\left(2A+B\right)+C\left(A+2B\right) \right] \right. \nonumber\\
&\left.+A+2B+C\right\}f_{00},\\
s_{11}^{(5)} &= ~\left(1-A\right) \left\{ \gamma \left[\alpha_5  B^2 C +\alpha_4 B\left(B+2C\right) \right.\right.\nonumber\\
&\left.\left.+ \alpha_3  \left(2B+C\right) +1 \right]+ \alpha_4 B^2 C \right. \nonumber\\
& \left.  +\alpha_3 B\left(B+2C\right) +2B+C \right\} f_{11},\\
s_{22}^{(5)} &= ~ \left(1-B\right) \left\{ \gamma \left[ \alpha_5  A BC  +\alpha_4 \left(AB+AC+BC\right) \right.\right.\nonumber\\
& \left.\left.+ \alpha_3 \left(A+B+C\right) +1 \right]+\alpha_4 ABC\right. \nonumber\\
&\left. +\alpha_3 \left(AB+AC+BC\right) +A+B+C\right\} f_{22}, \\
s_{33}^{(5)} &= ~s_{22}^{(5)} ,\\
s_{44}^{(5)} &= ~\left(1-C\right)\left\{ \gamma \left[ \alpha_5 AB^2 +\alpha_4 B \left(2A+B\right) \right.\right.\nonumber\\
& \left.\left. +\alpha_3 \left(A+2B\right) +1 \right] +\alpha_4 AB^2 \right. \nonumber\\
&\left.  + \alpha_3 B\left(2A+B\right) +A+2B\right\}f_{44}.
\end{align}

It is noted that besides the field equations for the physical and reference metrics, there exist some constraint equations associated with the Bianchi identities, which should hold in both $g$- and $f$-sectors. In particular, we will have the following constraint equations for $g_{\mu\nu}$ and $f_{\mu\nu}$ coming from the Bianchi identities shown in  Eqs. (\ref{constraint1}) and (\ref{constraint2}) as follows
\begin{align} \label{BI-constraint-1}
g^{00}\partial_0 {\cal H}^{(5)}_{00} &=  g^{11} \left[\Gamma^{0}_{11}(g){\cal H}_{00}^{(5)}+\Gamma^{1}_{10}(g){\cal H}_{11}^{(5)}\right] \nonumber\\
&+ 2g^{22} \left[\Gamma^{0}_{22}(g){\cal H}_{00}^{(5)}+\Gamma^{2}_{20}(g){\cal H}_{22}^{(5)}\right] \nonumber\\
&+g^{44} \left[\Gamma^{0}_{44}(g){\cal H}_{00}^{(5)}+\Gamma^{4}_{40}(g){\cal H}_{44}^{(5)}\right]
\end{align}
and
\begin{align}\label{BI-constraint-2}
& f^{00} \left\{ \partial_0 \left[ \frac{\sqrt{g}}{\sqrt{f}}s_{00}^{(5)}\right] -2 \frac{\sqrt{g}}{\sqrt{f}}\Gamma^{0}_{00}(f) s_{00}^{(5)} \right\} \nonumber\\
&= \frac{\sqrt{g}}{\sqrt{f}}f^{11} \left[ \Gamma^{0}_{11}(f)s_{00}^{(5)} +\Gamma^{1}_{10}(f) s_{11}^{(5)}\right] \nonumber\\
&+ 2\frac{\sqrt{g}}{\sqrt{f}}f^{22} \left[ \Gamma^{0}_{22}(f)s_{00}^{(5)} +\Gamma^{2}_{20}(f) s_{22}^{(5)}\right] \nonumber\\
&+ \frac{\sqrt{g}}{\sqrt{f}}f^{44} \left[ \Gamma^{0}_{44}(f)s_{00}^{(5)} +\Gamma^{4}_{40}(f) s_{44}^{(5)}\right] ,
\end{align}
 where we have set $N_1(t) =1$, i.e., $\gamma(t) =1- N_2(t)$ for convenience. Hence $\Gamma^{0}_{00}(g)=0$, while $\Gamma^{0}_{00}(f) =-\dot\gamma/(1-\gamma) \neq 0$.
Similar to the isotropic FLRW case, we will focus on a simple scenario by assuming that the right-hand sides of Eqs. (\ref{BI-constraint-1}) and (\ref{BI-constraint-2}) vanish altogether. As a result, this assumption can be done if we set
\begin{align}
\Gamma^{0}_{11}(g){\cal H}_{00}^{(5)}+\Gamma^{1}_{10}(g){\cal H}_{11}^{(5)} =&~\Gamma^{0}_{22}(g){\cal H}_{00}^{(5)}+\Gamma^{2}_{20}(g){\cal H}_{22}^{(5)} \nonumber\\
=&~\Gamma^{0}_{44}(g){\cal H}_{00}^{(5)}+\Gamma^{4}_{40}(g){\cal H}_{44}^{(5)} \nonumber\\
=&~0
\end{align}
along with
\begin{align}
\Gamma^{0}_{11}(f)s_{00}^{(5)} +\Gamma^{1}_{10}(f) s_{11}^{(5)} =&~ \Gamma^{0}_{22}(f)s_{00}^{(5)} +\Gamma^{2}_{20}(f) s_{22}^{(5)} \nonumber\\
=&~\Gamma^{0}_{44}(f)s_{00}^{(5)} +\Gamma^{4}_{40}(f) s_{44}^{(5)}\nonumber\\
=&~0,
\end{align}
which will  easily be fulfilled with the following simple solution: 
\begin{equation}
\gamma =A=B=C .
\end{equation}  
Furthermore,  Eq. (\ref{BI-constraint-1}) can be reduced, under this assumption, to
\begin{equation}
\partial_0 {\cal H}^{(5)}_{00} =0,
\end{equation}
which is equivalent to
\begin{equation}
{\cal H}^{(5)}_{00} = A\left(\sigma A^3 +4 \beta A^2 +6\alpha A +4\right) = \text{constant}.
\end{equation}

Since $\alpha$, $\beta$, and $\sigma$ are all constant coefficients, $A$ must be constant too. Hence,  $\gamma$, $B$, and $C$ will also act as  constants since they are all  equal to $A$ as mentioned above, resulting that the reference metric $f_{\mu\nu}$ will be proportional to the physical metric $g_{\mu\nu}$, i.e.,
\begin{equation}
 f_{\mu\nu} =(1- {\tilde C})^2 g_{\mu\nu},
\end{equation}
 where $ \tilde C$ is a constant, which should be real definite for expanding universes. In particular, we can set that
\begin{equation}
\gamma =A=B=C=\tilde C .
\end{equation}
As a result, this solution implies that
\begin{equation}
\eta = 1 \Rightarrow \sigma_1 =\sigma_2.
\end{equation}
This result indicates that both physical and reference metrics share the same anisotropic deviation. Note that we do not have the similar result for the other scale factors, $\alpha_i$ and $\sigma_i$. However, their time derivatives obey the same thing due to the fact that $\dot \gamma = \dot A =\dot B = \dot C =0$, i.e.,
\begin{equation}
\dot\alpha_1 =\dot\alpha_2, ~ \dot\beta_1 =\dot\beta_2, ~\ddot\alpha_1 =\ddot\alpha_2, ~\ddot\beta_1 =\ddot\beta_2 .
\end{equation}

Thanks to these results, we are able to write down explicitly the Einstein equations for the physical metric shown in Eq. (\ref{Einstein-5d}) as follows
\begin{align} \label{physical-Einstein-00}
 3  \left(\dot\alpha_1^2  -\dot\sigma_1^2+\dot\alpha_1 \dot\beta_1\right)  =&~\tilde \Lambda_0^g,\\
\label{physical-Einstein-11}
 2\ddot\alpha_1+ \ddot\beta_1+ 2\ddot\sigma_1  +3\dot\alpha_1^2 +\dot\beta_1^2+3\dot\sigma_1^2 & \nonumber\\
+ 2 \left(\dot\alpha_1 \dot\beta_1 + 3\dot\alpha_1 \dot\sigma_1 + \dot\beta_1 \dot\sigma_1 \right) =&~ \tilde \Lambda_0^g, \\
\label{physical-Einstein-22}
  2\ddot\alpha_1 + \ddot\beta_1- \ddot\sigma_1  +3\dot\alpha_1^2 +\dot\beta_1^2+3\dot\sigma_1^2 & \nonumber\\
+  \left(2\dot\alpha_1 \dot\beta_1 - 3\dot\alpha_1 \dot\sigma_1 - \dot\beta_1 \dot\sigma_1 \right) =&~ \tilde \Lambda_0^g,\\
\label{physical-Einstein-44}
 3  \left(\ddot\alpha_1 +2\dot\alpha_1^2+\dot\sigma_1^2\right) =&~\tilde \Lambda_0^g,
\end{align}
where an effective cosmological constant in the $g$-sector, $\tilde \Lambda_0^g$, is defined by
\begin{equation}
\tilde \Lambda_0^g = - \frac{\tilde C}{\tilde M_g^2} \left( \sigma \tilde C^3 +4\beta \tilde C^2 +6\alpha \tilde C +4 \right) ,
\end{equation}
with $\tilde M_g^2 = M_g^2/(m^2 M_{\text{eff}}^2)$.
Moreover, Eqs. (\ref{physical-Einstein-11}) and (\ref{physical-Einstein-22}) can be further reduced to 
\begin{align}
\label{physical-Einstein-11-m}
\ddot\sigma_1 +\dot\sigma_1 \left(3\dot\alpha_1 +\dot\beta_1\right) &=0,\\
\label{physical-Einstein-22-m}
 \ddot\beta_1  -2\dot\alpha_1^2 +2\dot\sigma_1^2 +\dot\beta_1^2 +\dot\alpha_1 \dot\beta_1 &=0,
\end{align}
with the help of Eqs. (\ref{physical-Einstein-00}) and (\ref{physical-Einstein-44}).

Similarly, we have the following field equations from the Einstein equations (\ref{5d-equation-f}) for the dynamical reference metric $f_{\mu\nu}$:
\begin{align} \label{reference-Einstein-00}
 3  \left(\dot\alpha_2^2  -\dot\sigma_2^2+\dot\alpha_2 \dot\beta_2 \right) =&~(1-\tilde C)^2 \tilde \Lambda_0^f,\\
\label{reference-Einstein-11}
 2\ddot\alpha_2+ \ddot\beta_2+ 2\ddot\sigma_2  +3\dot\alpha_2^2 +\dot\beta_2^2+3\dot\sigma_2^2  & \nonumber\\
+ 2 \left(\dot\alpha_2 \dot\beta_2 + 3\dot\alpha_2 \dot\sigma_2 + \dot\beta_2 \dot\sigma_2 \right)=&~(1-\tilde C)^2 \tilde \Lambda_0^f ,\\ 
\label{reference-Einstein-22}
 2\ddot\alpha_2 + \ddot\beta_2- \ddot\sigma_2  +3\dot\alpha_2^2 +\dot\beta_2^2+3\dot\sigma_2^2  & \nonumber\\
+  \left(2\dot\alpha_2 \dot\beta_2 - 3\dot\alpha_2 \dot\sigma_2 - \dot\beta_2 \dot\sigma_2 \right)=&~(1-\tilde C)^2 \tilde \Lambda_0^f, \\
\label{reference-Einstein-44}
3  \left(\ddot\alpha_2 +2\dot\alpha_2^2+\dot\sigma_2^2\right)  =&~(1-\tilde C)^2 \tilde \Lambda_0^f,
\end{align}
where the expression of an effective cosmological constant  $\tilde \Lambda_0^f$ in the $f$-sector is given by
\begin{equation}
\tilde \Lambda_0^f =\frac{\tilde C}{\tilde M_f^2 (1-\tilde C)^4} \left(\alpha_5 \tilde C^3 +4\alpha_4 \tilde C^2 +6 \alpha_3 \tilde C +4\right) ,
\end{equation}
 with $\tilde M_f^2 = M_f^2/(m^2 M_{\text{eff}}^2)$ as defined in the previous section for convenience. Additionally, the solution $\dot\gamma=0$ has been used in order to derive the above equations for $f_{\mu\nu}$. 
Moreover, we can obtain simplified equations from Eqs. (\ref{reference-Einstein-11}) and (\ref{reference-Einstein-22}):
\begin{align}
\label{reference-Einstein-11-m}
\ddot\sigma_2 +\dot\sigma_2 \left(3\dot\alpha_2 +\dot\beta_2\right) &=0,\\
\label{reference-Einstein-22-m}
 \ddot\beta_2  -2\dot\alpha_2^2 +2\dot\sigma_2^2 +\dot\beta_2^2 +\dot\alpha_2 \dot\beta_2 &=0,
\end{align}
with the help of the other ones, Eqs. (\ref{reference-Einstein-00}) and (\ref{reference-Einstein-44}). 

Now, by noting the result $\dot\alpha_1=\dot\alpha_2$, $\ddot\alpha_1 =\ddot\alpha_2$, $\dot\sigma_1 =\dot\sigma_2$, $\ddot\sigma_1 =\ddot\sigma_2$, $\dot\beta_1 =\dot\beta_2$, and $\ddot\beta_1 =\ddot\beta_2$, we come to a conclusion that 
\begin{equation}
\tilde \Lambda_0^g=(1-\tilde C)^2 \tilde \Lambda_0^f,
\end{equation}
 which implies an equation of $\tilde C$:
\begin{eqnarray} \label{eq-of-tilde-C}
&&\sigma \tilde C^5 -2 \left(\sigma-2\beta \right) \tilde C^4 + \left(\sigma-8\beta + 6\alpha + \alpha_5 \tilde M^2 \right)\tilde C^3 \nonumber\\
&&+ 4\left(\beta -3\alpha +\alpha_4 \tilde M^2+1\right)\tilde C^2 \nonumber\\ 
&&+2\left(3\alpha+3\alpha_3\tilde M^2-4\right) \tilde C +4 \left( \tilde M^2+1 \right) =0,
\end{eqnarray}
with the dimensionless parameter $\tilde M^2 \equiv \tilde M_g^2 / \tilde M_f^2$ as defined in the FLRW case. This equation turns out to be identical to  Eq. (\ref{eq-of-hat-C}) of $\hat C$ for the FLRW metric.  Hence,  $\tilde \Lambda_0^g$  will be equal to $\hat\Lambda_0^g$ for the FLRW metric if $\tilde C =\hat C$. In addition, it appears   in the isotropic limit corresponding the setting $ \sigma_i=0$ and $\beta_i =\alpha_i$ that all above field equations will be identical to that found in the previous section for the FLRW metric.

In conclusion, we have ended up with two sets of differential field equations: (i) Eqs. (\ref{physical-Einstein-00}), (\ref{physical-Einstein-44}), (\ref{physical-Einstein-11-m}), and (\ref{physical-Einstein-22-m}) for the $g$-sector of metric $g_{\mu\nu}$ and (ii) Eqs. (\ref{reference-Einstein-00}), (\ref{reference-Einstein-44}), (\ref{reference-Einstein-11-m}), and (\ref{reference-Einstein-22-m}) for the $f$-sector of metric $f_{\mu\nu}$. As a result,  two sets become identical to each other due to the solution that $f_{\mu\nu} =(1- \tilde C)^2 g_{\mu\nu}$, where the value of the constant $\tilde C$ has been determined by Eq.  (\ref{eq-of-tilde-C}). Therefore, we only need to solve one of these two sets of field equations for the scale factors, which might describe the evolution of our current universe.
\subsection{Analytical solutions}
It turns out that the field equations (\ref{physical-Einstein-00}), (\ref{physical-Einstein-44}), (\ref{physical-Einstein-11-m}), and (\ref{physical-Einstein-22-m}) look similar to that investigated in the five-dimensional dRGT theory with the five-dimensional Bianchi type I metrics \cite{TQD}. Hence, we will employ the method used in {Refs.~}\cite{WFK,TQD} to solve one of two sets of differential field equations shown above for the Bianchi type I metrics. As a result, we can obtain from Eqs. (\ref{physical-Einstein-00}), (\ref{physical-Einstein-44}), and (\ref{physical-Einstein-22-m}) two equations of two scale factors of physical metric:
\begin{align}
3 \ddot\beta_1 +3\dot\beta_1^2 +9\dot\alpha_1 \dot\beta_1 &= 2 \tilde \Lambda_0^g,\\
6\ddot\alpha_1 -3\ddot\beta_1 +18\dot\alpha_1^2 -3\dot\alpha_1 \dot\beta_1 -3\dot\beta_1^2 &=2 \tilde \Lambda_0^g,
\end{align}
which can be used to deduce a helpful relation: 
\begin{equation} \label{useful relation}
18 \left(\ddot\alpha_1 +3\dot\alpha_1^2\right) -6\left(\ddot\beta_1 +\dot\beta_1^2\right) = 8 \tilde \Lambda_0^g.
\end{equation}
Now, we introduce additional variables such as
\begin{equation}
 V_1 = \exp [3\alpha_1];~ V_2 =\exp[\beta_1] ,
\end{equation}
 as used in {Refs.~}\cite{WFK,TQD}. As a result, this introducing will leads Eq. (\ref{useful relation}) to 
\begin{equation}
\frac{\ddot V_1}{V_1} -\frac{\ddot V_2}{V_2} =\frac{4\tilde \Lambda_0^g}{3},
\end{equation}
which will be reduced to 
\begin{align} \label{equation of V1}
\ddot V_1 &=9 \tilde H_1^2 V_1,\\
\label{equation of V2}
\ddot V_2 &=9 \bar H_1^2 V_2,
\end{align}
if we assume that
\begin{equation}
\frac{\ddot V_2}{V_2} = V_0^g \frac{\ddot V_1}{V_1},
\end{equation}
where $V_0^g$ is a constant, $\tilde H_1^2 = 4\tilde \Lambda_0^g/27(1-V_0^g)$ and $\bar H_1^2= V_0^g \tilde H_1^2$ with a requirement that $0<V_0^g<1$ due to the positivity of $\tilde H_1^2$. Furthermore, solving Eqs. (\ref{equation of V1}) and (\ref{equation of V2}) gives us analytic solutions for $V_1$ and $V_2$ \cite{WFK,TQD}:
\begin{align}
V_1 \equiv \exp[3\alpha_1] =& \exp[3\alpha_{01}] \nonumber\\
&\times \left[\cosh\left(3\tilde H_1 t\right)+\frac{\dot\alpha_{01}}{\tilde H_1}\sinh \left(3\tilde H_1 t\right)\right],\\
V_2 \equiv \exp[\beta_1] =&\exp[\beta_{01}] \nonumber\\
&\times \left[\cosh\left(3\bar H_1 t\right)+\frac{\dot\beta_{01}}{3\bar H_1}\sinh \left(3\bar H_1 t\right)\right],
\end{align}
where $\alpha_{01} \equiv \alpha_1 (t=0)$, $\dot\alpha_{01} \equiv \dot\alpha_1(t=0)$, $\beta_{01} \equiv \beta_1 (t=0)$, and $\dot\beta_{01} \equiv \dot\beta_1(t=0)$ are initial values. Thanks to these explicit solutions, we now define the value of the last scale factor of the physical metric $\sigma_1$ by integrating out Eq. (\ref{physical-Einstein-11-m}) to be
\begin{equation} \label{solution-sigma-1}
\dot\sigma_1 = k \exp[-3\alpha_1 -\beta_1],
\end{equation}
where $k$ is an integration constant. And integrating out this equation leads to its solution:
\begin{align}\label{physical-sigma1-solution}
\sigma_1 =&~\sigma_{01} +\sqrt{\dot\alpha_{01}^2+\dot\alpha_{01} \dot\beta_{01} -\frac{\tilde\Lambda_0^g}{3}} \nonumber\\
& \times \int \Biggl\{ \biggl[\cosh \left(3\tilde H_1 t\right)+\frac{\dot\alpha_{01}}{\tilde H_1} \sinh \left(3\tilde H_1 t \right) \biggr] \nonumber\\
& \times \biggl[\cosh \left(3\bar H_1 t\right)+\frac{\dot\beta_{01}}{3\bar H_1} \sinh \left(3\bar H_1 t \right) \biggr]\Biggr\}^{-1} dt,
\end{align}
here $\sigma_{01} \equiv \sigma_1 (t=0)$ acts as an initial value. In addition, we have used the initial condition coming from the Friedmann equation (\ref{physical-Einstein-00}):
\begin{equation}
\dot\alpha_{01}^2 +\dot\alpha_{01}\dot\beta_{01} - \frac{\tilde\Lambda_0^g}{3} = k^2 \exp\left[-6\alpha_{01} -2\beta_{01}\right]
\end{equation}
in order to derive the above solution of $\sigma_1$. 

For  the reference metric $f_{\mu\nu}$, we obtain the corresponding solutions for its scale factors as follows
\begin{align}
 \exp[\alpha_2] &= (1-{\tilde C}) \exp[\alpha_1] ,\\
 \exp[\beta_2] &=(1-{\tilde C}) \exp[\beta_1] ,\\
\sigma_2 &= \sigma_1,
\end{align}
since it has been assumed to be proportional to the physical metric, i.e., $f_{\mu\nu}=(1-\tilde C)^2 g_{\mu\nu}$, in order to satisfy the Bianchi identities. 

In conclusion, we have derived analytical Bianchi type I solutions for  the corresponding differential field equations: (i) Eqs. (\ref{physical-Einstein-00}), (\ref{physical-Einstein-44}), (\ref{physical-Einstein-11-m}), and (\ref{physical-Einstein-22-m}) for the $g$-sector and (ii) Eqs. (\ref{reference-Einstein-00}), (\ref{reference-Einstein-44}), (\ref{reference-Einstein-11-m}), and (\ref{reference-Einstein-22-m}) for the $f$-sector, as promised.  
\subsection{Stability analysis}
Given the above analytical anisotropic solutions of the physical metric, we  would like to study their stability to see whether they respect the well-known cosmic no-hair conjecture proposed by Hawking and his colleagues long time ago \cite{Hawking,counter-example,WFK1}. In particular, this conjecture postulates that the final state of the universe must be isotropic. In other words, if this conjecture holds then any anisotropic cosmological solution describing either early or current universes must be unstable and then decay to an isotropic state at late time.  However, it is worth noting that a complete proof for this conjecture has not been done up to now. A partial proof for Bianchi spaces by Wald can be found in {Ref.~}\cite{Hawking}. The proof of Wald has only deals with strong and dominant energy conditions without explicit perturbation analysis. Hence, in some models, in which the energy conditions do not hold clearly, this proof might not gives us valid conclusions on the validity of the no-hair conjecture. It turns out that  stability analysis based on perturbation approaches for anisotropic solutions should be performed, even when the energy conditions hold explicitly, in order to get correct conclusions of the fate of the no-hair conjecture. For example, it has been shown by perturbation analysis in {Refs.~}\cite{counter-example,WFK1} that the no-hair conjecture turns out to be violated in  supergravity-motivated models, where a scalar field, either canonical or non-canonical, is coupled with a $U(1)$ field.  As a result, these models have admitted  Bianchi type I inflationary solutions, which have been shown to be stable against field perturbations.  Besides this scalar-vector model, the no-hair conjecture has also faced other counter-examples in the context of the massive gravity as investigated in \cite{TQD,WFK}. Note also that the cosmic no-hair conjecture has been examined in the context of four-dimensional bigravity in \cite{soda}. As a result, the cosmic no-hair conjecture seems to be valid for de-Sitter spacetimes in the bigravity \cite{soda}.

Hence, we now would like to perturb Eqs.  (\ref{physical-Einstein-00}), (\ref{physical-Einstein-11-m}), and (\ref{physical-Einstein-22-m}) by
taking exponential perturbations: $\delta \alpha_1 = C_{\alpha_1} \exp [\omega t]$, $\delta \sigma_1 = C_{\sigma_1} \exp [\omega t]$ and $\delta \beta_1 = C_{\beta_1} \exp [\omega t]$. Consequently, we will have the following perturbation equations, which can be written as a matrix equation:
\begin{equation}\label{matrix}
{\cal D} \left( {\begin{array}{*{20}c}
    {\cal C}_{\alpha_1} \\
    {\cal C}_{\sigma_1}   \\
    {\cal C}_{\beta_1} \\
 \end{array} } \right) \equiv \left[ {\begin{array}{*{20}c}
   {A_{11} } & {A_{12} } & {A_{13} }  \\
   {A_{21} } & {A_{22} } & {A_{23} } \\
   {A_{31} } & {A_{32} } & {A_{33} } \\
 \end{array} } \right]\left( {\begin{array}{*{20}c}
    {\cal C}_{\alpha_1} \\
    {\cal C}_{\sigma_1}   \\
    {\cal C}_{\beta_1} \\
 \end{array} } \right) = 0,
\end{equation}
where
\begin{align}
&A_{11} = \left(2\dot\alpha_1 +\dot\beta_1 \right)\omega; ~ A_{12} = -2\dot\sigma_1 \omega; ~ A_{13} = \dot\alpha_1 \omega; \\
&A_{21}= 3\dot\sigma_1 \omega; ~ A_{22}= \omega^2 +\left(3\dot\alpha_1 +\dot\beta_1\right)\omega; ~A_{23}=\dot\sigma_1 \omega; \\
&A_{31} = -\left(4\dot\alpha_1 -\dot\beta_1\right)\omega;~A_{32}= 4\dot\sigma_1 \omega ; 
\end{align}
\begin{align}
A_{33} = \omega^2 +\left(\dot\alpha_1 +2\dot\beta_1\right)\omega.
\end{align}
Mathematically, Eq. (\ref{matrix}) will admit non-trivial solutions only when
\begin{equation}
\det {\cal D}  = 0.
\end{equation}
It turns out that the determinant equation, $\det{\cal D}=0$,  can be rewritten as an equation of $\omega$:
\begin{equation} \label{omega-equation}
A_1 \omega^2 +B_1 \omega +C_1 =0,
\end{equation}
where 
\begin{align}
A_1 =&~2\dot\alpha_1 +\dot\beta_1,\\
B_1=&~3 \left[\dot\alpha \left(4\dot\alpha_1 +3 \dot\beta_1 \right) +\dot\beta_1^2 +2\dot\sigma_1^2 \right],\\
C_1=&~18\dot\alpha_1^2 \left(\dot\alpha_1 +\dot\beta_1\right)+2\dot\beta_1^2 \left(5\dot\alpha_1 +\dot\beta_1 \right) \nonumber\\
&+6\dot\sigma_1^2\left(3\dot\alpha_1 +\dot\beta_1\right).
\end{align}

Note that we have ignored the trivial solution, $\omega=0$, of this determinant equation. Now, we would like to point out whether the quadratic equation of $\omega$ shown above admits only non-positive roots by observing that if all coefficients $A_1$, $B_1$, and $C_1$ act as non-negative parameters, then Eq. (\ref{omega-equation}) will no longer admit any non-positive root. As a result, for expanding universes with $\dot\alpha_1 >0$ and $\dot\alpha_1 > \dot\sigma_1,~ \dot\beta_1$ then the above requirement can  easily be satisfied, meaning that the anisotropically expanding universes in the five-dimensional massive gravity with small spatial anisotropies are indeed stable against field perturbations. It is noted that we have obtained  four- and five-dimensional stable Bianchi type I solutions for the corresponding four- and five-dimensional massive gravity models in {Refs.~}\cite{TQD,WFK}. The result in this section provides one more counter-example to the cosmic no-hair conjecture  \cite{Hawking,counter-example,WFK1}.
\section{Schwarzschild-Tangherlini-(A)dS black holes} \label{sec6}
\subsection{Field equations}
In the previous sections, we have studied the five-dimensional massive bigravity for both homogeneous metrics, the isotropic FLRW metric and the  anisotropic Bianchi type I  metric. As a result, we have shown that a simple choice to make the total graviton term ${\cal U}_M$  constant is choosing that the reference metrics are proportional to the physical metrics, i.e., $f_{\mu\nu}=(1- {\cal C})^2 g_{\mu\nu}$, where ${\cal C}$ is a constant: ${\cal C}=\hat C$ for the FLRW metric and ${\cal C} =\tilde C$ for the Bianchi type I metric. More interestingly, the following equations of $\hat C$ and $\tilde C$, which will give us the values of the corresponding constants, have been derived to be identical to each other as shown in Eqs. (\ref{eq-of-hat-C}) and (\ref{eq-of-tilde-C}). Hence, one could expect that these equations would be a unique equation for determining value of proportional constants between the reference and physical metrics even when the metrics are more complicated than the FLRW and Bianchi type I ones, e.g., the  Schwarzschild-Tangherlini black hole \cite{5d-sch,5d-sch-stability,5d-review}, which will be studied in the rest of this section.

Following our recent work \cite{TQD}, we would like to seek the Schwarzschild-Tangherlini black hole  for the five-dimensional massive bigravity by considering  spherically symmetric metrics for both $g$- and $f$-sectors:
\begin{eqnarray} \label{five-dim-metric}
 g^{\text {5d}}_{\mu\nu}dx^{\mu}dx^{\nu} &= & -N_1^2\left(r\right) dt^2 +\frac{dr^2}{F_1^2\left(r\right)}+\frac{r^2 d\Omega_3^2}{H_1^2\left(r\right)},\\
 f^{\text {5d}}_{\mu\nu}dx^{\mu}dx^{\nu}&= & -N_2^2\left(r\right) dt^2 +\frac{dr^2}{F_2^2\left(r\right)}+\frac{r^2 d\Omega_3^2}{H_2^2\left(r\right)}, 
\end{eqnarray}
with
\begin{equation}
d\Omega_3^2=d\theta^2 +\sin^2 \theta d\varphi^2 + \sin^2 \theta \sin^2 \varphi d \psi^2.
\end{equation}
Here $N_i(r)$, $F_i(r)$, and $H_i(r)$ ($i=1-2$) are arbitrary functions of the radial coordinate $r$.  In addition, $(\theta,\varphi,\psi)$ with allowed ranges: $0\leq \theta \leq \pi$, $0\leq \varphi \leq \pi$, and $0\leq \psi \leq 2\pi$  are the spherical coordinates. Note that according to the  discussion in {Ref.~}\cite{TQD} we have set the non-diagonal element of metrics  associated with the non-diagonal components $g_{0r}$ (and $f_{0r}$)  to be zero due to the following constraint:
\begin{equation}
g_{0r}R_{00}-g_{00}R_{0r}=0,
\end{equation}
for simplicity.
Of course, one might seek non-diagonal solutions for the five-dimensional bigravity as investigations  for the four-dimensional massive gravity in {Refs.~}\cite{blackholes,blackholes-stability}. 

For convenience, let us recall some useful definitions defined in {Ref.~}\cite{TQD}. In particular, the non-vanishing components of ${\cal K}^{\mu}{ }_\nu$ turn out to be
\begin{eqnarray} \label{component-K}
{\cal K}^0{ }_0(r) &=&1- \frac{N_2}{N_1}, ~ {\cal K}^1{ }_1(r) = 1- \frac{F_1}{F_2},  \nonumber\\ 
 {\cal K}^2{ }_2(r)&= &{\cal K}^3{ }_3(r) = {\cal K}^4{ }_4(r)=1-\frac{H_1}{H_2}.
\end{eqnarray}
On the other hand, these non-vanishing components contribute to the massive graviton terms as follows:  
\begin{align}
{\cal U}_2 =&~ 2 \Bigl[ {\cal K}^0{ }_0 \left({\cal K}^1{ }_1 +3 {\cal K}^2{ }_2\right)+3 {\cal K}^2{ }_2 \left({\cal K}^1{ }_1 +{\cal K}^2{ }_2\right)   \Bigr], \\
{\cal U}_3 =& ~2 {\cal K}^2{ }_2 \Bigl[  3{\cal K}^0{ }_0 \left({\cal K}^1{ }_1 + {\cal K}^2{ }_2\right)+{\cal K}^2{ }_2 \left(3{\cal K}^1{ }_1 +{\cal K}^2{ }_2 \right) \Bigr],\\
{\cal U}_4 =& ~2\left({\cal K}^2{ }_2\right)^2 \Bigl[{\cal K}^0{ }_0 \left( 3{\cal K}^1{ }_1 + {\cal K}^2{ }_2\right) +{\cal K}^1{ }_1{\cal K}^2{ }_2     \Bigr], \\
{\cal U}_5 =&~2 {\cal K}^0{ }_0 {\cal K}^1{ }_1 \left({\cal K}^2{ }_2\right)^3 .
\end{align}
Hence, the total massive graviton term ${\cal U}_M \equiv {\cal U}_2 +\alpha_3 {\cal U}_3 +\alpha_4 {\cal U}_4 +\alpha_5 {\cal U}_5$ turns out to be
\begin{align} \label{Lagra-reduced}
{\cal U}_M &=~2\left\{ {\cal K}^0{ }_0 {\cal K}^1{ }_1\left[\alpha_5 \left({\cal K}^2{ }_2\right)^3+ 3\alpha_4 \left({\cal K}^2{ }_2\right)^2 +3 \alpha_3 {\cal K}^2{ }_2 +1\right]   \right.\nonumber\\ 
& \left.+{\cal K}^2{ }_2 \left[ \alpha_4 \left({\cal K}^2{ }_2\right)^2+3\alpha_3 {\cal K}^2{ }_2 +3\right] \left({\cal K}^0{ }_0 + {\cal K}^1{ }_1\right) \right. \nonumber\\
& \left. + \left({\cal K}^2{ }_2\right)^2 \left(\alpha_3{\cal K}^2{ }_2+3 \right)\right\}.
\end{align}

Similar to the previous sections, we are going to determine the non-vanishing components of the tensor ${\cal H}^{(5)}_{\mu\nu}$ defined in Eq. (\ref{def-of-H}). As a result, they turn out to be
\begin{align}
{\cal H}^{(5)}_{00} & = -\Bigl[\sigma {\cal K}^1{ }_1 \left({\cal K}^2{ }_2\right)^3 +\beta \left({\cal K}^2{ }_2\right)^2 \left(3{\cal K}^1{ }_1 +{\cal K}^2{ }_2\right) \nonumber\\
&+3\alpha {\cal K}^2{ }_2 \left({\cal K}^1{ }_1+{\cal K}^2{ }_2\right) + {\cal K}^1{ }_1 +3{\cal K}^2{ }_2\Bigl] g_{00},\\
{\cal H}^{(5)}_{11}& =-\Bigl[\sigma {\cal K}^0{ }_0 \left({\cal K}^2{ }_2\right)^3 +\beta \left({\cal K}^2{ }_2\right)^2 \left(3{\cal K}^0{ }_0 +{\cal K}^2{ }_2\right) \nonumber\\
&+3\alpha {\cal K}^2{ }_2 \left({\cal K}^0{ }_0+{\cal K}^2{ }_2\right) + {\cal K}^0{ }_0 +3{\cal K}^2{ }_2\Bigl] g_{11},\\
{\cal H}^{(5)}_{22}& =- \Bigl\{\sigma  {\cal K}^0{ }_0  {\cal K}^1{ }_1 \left({\cal K}^2{ }_2\right)^2 \nonumber\\
&+\beta {\cal K}^2{ }_2 \left(2{\cal K}^0{ }_0 {\cal K}^1{ }_1 +{\cal K}^0{ }_0{\cal K}^2{ }_2 +{\cal K}^1{ }_1{\cal K}^2{ }_2\right)  \nonumber\\
& +\alpha \left[{\cal K}^0{ }_0{\cal K}^1{ }_1+{\cal K}^2{ }_2 \left(2{\cal K}^0{ }_0+2{\cal K}^1{ }_1+{\cal K}^2{ }_2\right)  \right] \nonumber\\
&+ {\cal K}^0{ }_0+{\cal K}^1{ }_1+2{\cal K}^2{ }_2  \Bigl\} g_{22}, \\
{\cal H}^{(5)}_{44}&=~{\cal H}^{(5)}_{33}={\cal H}^{(5)}_{22}.
\end{align}
It is noted that along with the tensor ${\cal H}^{(5)}_{\mu\nu}$ for the physical metric $g_{\mu\nu}$, there exists the tensor $s^{(5)}_{\mu\nu}$ for the reference metric $f_{\mu\nu}$, whose non-vanishing components read
\begin{align}
s^{(5)}_{00} &=~ \left(1-{\cal K}^0{ }_0\right) \Bigl[\alpha_5 {\cal K}^1{ }_1 \left({\cal K}^2{ }_2\right)^3 \nonumber\\
& +\alpha_4 \left({\cal K}^2{ }_2\right)^2  \left(3{\cal K}^1{ }_1+{\cal K}^2{ }_2\right) +3\alpha_3 {\cal K}^2{ }_2 \left({\cal K}^1{ }_1 +{\cal K}^2{ }_2\right) \nonumber\\
& +{\cal K}^1{ }_1+3{\cal K}^2{ }_2 \Bigl]f_{00},\\
s^{(5)}_{11} &=~ \left(1-{\cal K}^1{ }_1\right) \Bigl[\alpha_5 {\cal K}^0{ }_0 \left({\cal K}^2{ }_2\right)^3 \nonumber\\
&+\alpha_4 \left({\cal K}^2{ }_2\right)^2  \left(3{\cal K}^0{ }_0+{\cal K}^2{ }_2\right) +3\alpha_3 {\cal K}^2{ }_2 \left({\cal K}^0{ }_0 +{\cal K}^2{ }_2\right) \nonumber\\
& +{\cal K}^0{ }_0+3{\cal K}^2{ }_2 \Bigl]f_{11},\\
s^{(5)}_{22} &=~ \left(1-{\cal K}^2{ }_2\right) \Bigl\{ \alpha_5 {\cal K}^0{ }_0  {\cal K}^1{ }_1 \left( {\cal K}^2{ }_2 \right)^2 \nonumber\\
&+ \alpha_4  {\cal K}^2{ }_2 \left(2 {\cal K}^0{ }_0  {\cal K}^1{ }_1 + {\cal K}^0{ }_0  {\cal K}^2{ }_2 + {\cal K}^1{ }_1  {\cal K}^2{ }_2    \right) \nonumber\\
& + \alpha_3 \left[ \left({\cal K}^2{ }_2\right)^2 +2  {\cal K}^2{ }_2 \left( {\cal K}^0{ }_0 + {\cal K}^1{ }_1 \right) + {\cal K}^0{ }_0  {\cal K}^1{ }_1 \right] \nonumber\\
&+  {\cal K}^0{ }_0 + {\cal K}^1{ }_1 +2  {\cal K}^2{ }_2 \Bigl\} f_{22},\\  
s^{(5)}_{44} &=~ s^{(5)}_{33} =s^{(5)}_{22}.
\end{align}
Armed with these explicit definitions, we will seek analytical solutions for both physical and reference field equations in the next subsection. 
\subsection{Analytical solutions}
Now, let us come back the Bianchi constraints for the physical and reference metrics as shown in Eqs. (\ref{constraint1}) and (\ref{constraint2}). As a result, Eq. (\ref{constraint1}) leads to a set of non-vanishing component equations:
\begin{align}\label{constraint-physical-Sch-1}
&g^{11}\left[\partial_r {\cal H}^{(5)}_{11}-2\Gamma^1_{11}(g){\cal H}^{(5)}_{11}\right] \nonumber\\
& = g^{00}\left[\Gamma^1_{00}(g) {\cal H}^{(5)}_{11} +\Gamma^0_{01}(g){\cal H}^{(5)}_{00} \right]\nonumber\\
&+g^{22} \left[\Gamma^1_{22}(g){\cal H}^{(5)}_{11} +\Gamma^2_{21}(g){\cal H}^{(5)}_{22} \right]\nonumber\\
&+g^{33} \left[\Gamma^1_{33}(g){\cal H}^{(5)}_{11} +\Gamma^3_{31}(g){\cal H}^{(5)}_{33} \right] \nonumber\\
&+g^{44} \left[\Gamma^1_{44}(g){\cal H}^{(5)}_{11} + \Gamma^4_{41}(g){\cal H}^{(5)}_{44} \right],
\end{align}
\begin{align}
\label{constraint-physical-Sch-2}
&g^{33} \left[\Gamma^2_{33}(g){\cal H}^{(5)}_{22} + \Gamma^3_{32}(g){\cal H}^{(5)}_{33}\right]  \nonumber\\
 &+ g^{44} \left[\Gamma^2_{44}(g){\cal H}^{(5)}_{22} + \Gamma^4_{42}(g){\cal H}^{(5)}_{44}\right]=0,
\end{align}
\begin{align}
\label{constraint-physical-Sch-3}
g^{44} \left[\Gamma^3_{44}(g){\cal H}^{(5)}_{33} + \Gamma^4_{43}(g){\cal H}^{(5)}_{44}\right] &=0.
\end{align}
On the other hand, Eq. (\ref{constraint2}) implies another set of non-vanishing component equations:
\begin{align}\label{constraint-reference-Sch-1}
&f^{11}\left\{ \partial_r \left[ \frac{\sqrt{g}}{\sqrt{f}}s^{(5)}_{11}\right]-2\frac{\sqrt{g}}{\sqrt{f}}\Gamma^1_{11}(f)s^{(5)}_{11}\right\} \nonumber\\
  &=\frac{\sqrt{g}}{\sqrt{f}} f^{00}\left[\Gamma^1_{00}(f) s^{(5)}_{11} +\Gamma^0_{01}(f)s^{(5)}_{00} \right] \nonumber\\
&+\frac{\sqrt{g}}{\sqrt{f}}f^{22} \left[\Gamma^1_{22}(f)s^{(5)}_{11} +\Gamma^2_{21}(f)s^{(5)}_{22} \right]\nonumber\\
&+\frac{\sqrt{g}}{\sqrt{f}}f^{33} \left[\Gamma^1_{33}(f)s^{(5)}_{11} +\Gamma^3_{31}(f)s^{(5)}_{33} \right] \nonumber\\
& +\frac{\sqrt{g}}{\sqrt{f}}f^{44} \left[\Gamma^1_{44}(f)s^{(5)}_{11} + \Gamma^4_{41}(f)s^{(5)}_{44} \right],
\end{align}
\begin{align}
\label{constraint-reference-Sch-2}
 & f^{33} \left[\Gamma^2_{33}(f)s^{(5)}_{22} + \Gamma^3_{32}(f)s^{(5)}_{33}\right] \nonumber\\
& + f^{44} \left[\Gamma^2_{44}(f)s^{(5)}_{22} + \Gamma^4_{42}(f)s^{(5)}_{44}\right] =0,
\end{align}
\begin{align}
\label{constraint-reference-Sch-3}
f^{44} \left[\Gamma^3_{44}(f)s^{(5)}_{33} + \Gamma^4_{43}(f)s^{(5)}_{44}\right] &=0.
\end{align}

Similar to the previous studies for the FLRW and Bianchi metrics, a simple solution to both Eqs. (\ref{constraint-physical-Sch-2}) and (\ref{constraint-physical-Sch-3}) can be solved to be
\begin{equation} \label{sch-solution-constraint}
 {\cal K}^2{ }_2={\cal K}^1{ }_1={\cal K}^0{ }_0 ,
\end{equation}
which also makes the right-hand side of Eq. (\ref{constraint-physical-Sch-1}) zero.  As a result,  the left-hand side of Eq. (\ref{constraint-physical-Sch-1}) now reduces to 
\begin{equation}
\partial_r \left[\sigma \left({\cal K}^0{ }_0\right)^4 +4\beta \left({\cal K}^0{ }_0\right)^3 +6\alpha \left({\cal K}^0{ }_0\right)^2 +4{\cal K}^0{ }_0 \right]=0.
\end{equation}
This equation can be integrated out to give an equation of ${\cal K}^0{ }_0$:
\begin{equation}
\sigma \left({\cal K}^0{ }_0\right)^4 +4\beta \left({\cal K}^0{ }_0\right)^3 +6\alpha \left({\cal K}^0{ }_0\right)^2 +4{\cal K}^0{ }_0 = \text{constant}.
\end{equation}
Hence, once this equation is solved the corresponding solutions  ${\cal K}^0{ }_0$ must be constant since all coefficients $\sigma$, $\beta$, and $\alpha$ are also constant. Therefore, we will set that
\begin{equation} \label{sch-solution-constraint}
 {\cal K}^2{ }_2={\cal K}^1{ }_1={\cal K}^0{ }_0 = \bar C ,
\end{equation}
where $\bar C$ is a constant, whose value will be figured out later. 
Furthermore, this solution is equivalent to the scenario  that the physical metric is proportional to the reference metric, 
\begin{equation}
f_{\mu\nu} =(1 -\bar C)^2 g_{\mu\nu},
\end{equation}
which has been discussed extensively for a number black hole solutions of the four-dimensional massive (bi)gravity, e.g., see some review papers in {Ref.~}\cite{blackholes-review}.
It is straightforward to check that the solution shown in Eq. (\ref{sch-solution-constraint}) is also a solution of Eqs. (\ref{constraint-reference-Sch-1}), (\ref{constraint-reference-Sch-2}), and (\ref{constraint-reference-Sch-3}) for the reference metric.

It is noted that the solution displayed in Eq. (\ref{sch-solution-constraint}) also implies that the massive graviton terms ${\cal U}_i$'s ($i=2-5$) along with the total graviton term ${\cal U}_M$ all turn out to be nothing but constants. In particular, we have the corresponding effective cosmological constant for the physical metric:
\begin{equation}
{\bar \Lambda}_0^g = -\frac{\bar C}{{\tilde M}^2_g} \left(\sigma \bar C^3+4\beta \bar C^2+6\alpha \bar C +4 \right)
\end{equation}
along with that for reference metric:
\begin{equation}
{\bar \Lambda}_0^f = \frac{\bar C}{{\tilde M}^2_f (1-\bar C)^4} \left(\alpha_5 \bar C^3+4\alpha_4 \bar C^2+6\alpha_3 \bar C +4 \right).
\end{equation}
Thanks to these results, the Einstein field equations for the $g$- and $f$-sectors now reduce to simple forms:
\begin{align} \label{sch-equation-g}
R_{\mu\nu}(g)-\frac{1}{2}g_{\mu\nu}R(g)+ \bar\Lambda_0^g g_{\mu\nu}&=0, \\
 \label{sch-equation-f}
R_{\mu\nu}(f)-\frac{1}{2}f_{\mu\nu}R(f)+ \bar\Lambda_0^f f_{\mu\nu} & =0.
\end{align}
As found in {Refs.~}\cite{5d-sch,5d-sch-stability}, the field Einstein equations (\ref{sch-equation-g}) with the effective cosmological constant $\bar\Lambda_0^g$ admit non-trivial solutions:
\begin{eqnarray}
N_1^2(r)&=&F_1^2(r)=f(r)=1-\frac{\mu}{r^2}-\frac{\bar\Lambda_0^g}{6}r^2,\\
H_1^2(r)&=&1.
\end{eqnarray}
which correspond to the following metric:
\begin{equation} \label{five-dim-metric-1}
g^{\text {5d}}_{\mu\nu}dx^{\mu}dx^{\nu} = -f(r)dt^2 +\frac{dr^2}{f(r)}+r^2d\Omega_3^2,
\end{equation}
where $\mu \equiv 8 G_5 M/(3\pi) $ is a mass parameter with $M$ and $G_5$ standing for the mass of source and  the five-dimensional Newton constant, respectively. Furthermore,  the metric shown in Eq. (\ref{five-dim-metric-1}) will be regarded as the  Schwarzschild-Tangherlini-de Sitter or Schwarzschild-Tangherlini-anti-de Sitter black hole if $\bar\Lambda_0^g >0$ or  $\bar\Lambda_0^g <0$, respectively. On the other hand, we will have the pure Schwarzschild-Tangherlini black hole for the case of vanishing $\bar\Lambda_0^g$. 

It is noted that we have worked on a specific scenario, where the reference metric is taken to be proportional to the physical metric such as $f_{\mu\nu}=(1-\bar C)^2 g_{\mu\nu}$, in order to satisfy the Bianchi constraints. Consequently, this choice leads to the constant-like behavior of the massive graviton terms in both $g$- and $f$-sectors as shown above. Therefore, it is straightforward to obtain the corresponding reference metric:
 \begin{equation} \label{five-dim-metric-1}
f^{\text {5d}}_{\mu\nu}dx^{\mu}dx^{\nu} =(1-\bar C)^2\left[ -f(r)dt^2 +\frac{dr^2}{f(r)}+r^2d\Omega_3^2\right],
\end{equation}
Here, the value of the constant $\bar C$ can be figured out from the constraint that the effective cosmological constant $\bar\Lambda_0^g$  must be equal to $(1-\bar C)^2\bar\Lambda_0^f$. As a result, this requirement implies  the following equation of $\bar C$:
\begin{eqnarray} \label{eq-of-bar-C}
&&\sigma \bar C^5 -2 \left(\sigma-2\beta \right) \bar C^4 + \left(\sigma-8\beta + 6\alpha + \alpha_5 \tilde M^2 \right)\bar C^3 \nonumber\\
&&+ 4\left(\beta -3\alpha +\alpha_4 \tilde M^2+1\right)\bar C^2 \nonumber\\ 
&&+2\left(3\alpha+3\alpha_3\tilde M^2-4\right) \bar C +4 \left( \tilde M^2+1 \right) =0,
\end{eqnarray}
which looks identical to that of $\hat C$ for the FLRW metric and that of $\tilde C$ for the Bianchi type I metric derived in the previous sections.

For the stability of the Schwarzschild-Tangherlini-(A)dS black holes in the context of five-dimensional massive bigravity, one might follow some recent investigations within  four-dimensional spacetimes, which have been done in {Ref.~}\cite{blackholes-stability}. In particular, the authors of papers in {Ref.~}\cite{blackholes-stability} have found by a systematic stability analysis that the  four-dimensional Schwarzschild black holes  turn out to be unstable against radial perturbations whatever the reference metric is dynamical or not. Therefore, the Schwarzschild-Tangherlini-(A)dS black holes of five- (or higher)-dimensional massive (bi)gravity theory might also be expected to be unstable. However, it is noted that  the existence of new graviton terms such as ${\cal U}_5$  might affect on the stability of black holes of higher dimensional massive (bi)gravity theory. Hence,  it addresses further investigations to obtain valid conclusions for the stability of the Schwarzschild-Tangherlini-(A)dS black holes in the context of five-dimensional massive bigravity. On the other hand, if one would like to have stable black holes to the five-dimensional massive (bi)gravity theory, one might think of  non-bidiagonal metrics according to the last paper in {Ref.~}\cite{blackholes-stability}, which has focused only on four-dimensional spacetimes.
\section{Four-dimensional limit} \label{sec7}
As a result, we have shown that one of choices to make the graviton terms constant is that the dynamical reference metric $f_{\mu\nu}$ is taken to be proportional to the physical one $g_{\mu\nu}$ under the relation $f_{\mu\nu}=(1-{\cal C})^2 g_{\mu\nu}$ with ${\cal C} =\hat C$, $\tilde C$, and $\bar C$ for the FLRW, Bianchi type I, and Schwarzschild-Tangherlini metrics, respectively. Moreover,  the equations of proportional factors, $\hat C$, $\tilde C$, and $\bar C$, which will give us their corresponding values,  have appeared to be in the same form as displayed in Eqs. (\ref{eq-of-hat-C}), (\ref{eq-of-tilde-C}), and (\ref{eq-of-bar-C}):
\begin{eqnarray} \label{eq-of-cal-C}
&&\sigma {\cal C}^5 -2 \left(\sigma-2\beta \right) {\cal C}^4 + \left(\sigma-8\beta + 6\alpha + \alpha_5 \tilde M^2 \right){\cal C}^3 \nonumber\\
&&+ 4\left(\beta -3\alpha +\alpha_4 \tilde M^2+1\right){\cal C}^2 \nonumber\\ 
&&+2\left(3\alpha+3\alpha_3\tilde M^2-4\right) {\cal C} +4 \left( \tilde M^2+1 \right) =0.
\end{eqnarray}
We might expect that this equation might be valid for a number of reference and physical metrics, which are diagonal and proportional to each other, of the five-dimensional massive bigravity. Once again, we note that in the dRGT limit, where $s^{(5)}_{\mu\nu}=0$ due to the non-dynamical property of the reference metric $f_{\mu\nu}$, the corresponding equation of ${\cal C}$ turns out to be identical to that investigated in \cite{TQD}:
\begin{equation}
\alpha_5 {\cal C}^3 +4\alpha_4 {\cal C}^2 +6\alpha_3 {\cal C}+4 =0.
\end{equation}

Note that the four-dimensional limit of the bigravity cannot be recovered by simply setting $\alpha_5=0$ in the five-dimensional scenario. The reason is that the graviton terms living in  five-dimensional spacetimes contain more terms than that in four-dimensional spacetimes.  Therefore, setting $\alpha_5 =0$ does not kill extra terms in ${\cal U}_i$'s. Following the investigations done in {Ref.~}\cite{TQD}, the coefficient $\alpha_5$ associated with the existence of ${\cal U}_5$ should be fine-tuned in order to recover effective cosmological constants of four-dimensional models. 

To improve this claim, we now consider an example, in which the four-dimensional FLRW metric is adopted for both physical and reference metrics of the following four-dimensional massive bigravity as follows
\begin{eqnarray} \label{4d-frw-1}
ds_{\text {4d}}^2 (g_{\mu\nu}) &=& -N_1^2(t)dt^2 + a_1^2(t) d\vec{x}^2, \\
 \label{4d-frw-2}
ds_{\text {4d}}^2 (f_{\mu\nu}) &=& -N_2^2(t)dt^2 + a_2^2(t)d\vec{x}^2.
\end{eqnarray}
For details of four-dimensional massive bigravity, one can see the section \ref{sec2}. 
As a result, the corresponding four-dimensional graviton terms ${\cal U}_i$'s ($i=2-4$) are defined to be
\begin{align}
{\cal U}_2 =&~ 3\Sigma \left(\gamma+\Sigma\right); ~{\cal U}_3 =\Sigma^2 \left(\gamma+\Sigma\right); ~
{\cal U}_4 =\gamma \Sigma^3; \\
{\cal U}_M =&~\Sigma \left[ \left(\alpha_4 \gamma +\alpha_3\right) \Sigma^2+3 \left(\alpha_3 \gamma+1\right)\Sigma+3\gamma \right].
\end{align}
Note again that ${\cal U}_{5} =0$ in all four-dimensional spacetimes. 
Additionally, the following components of tensor ${\cal H}_{\mu\nu}(g)$ appearing in the four-dimensional Einstein field equations for the physical metric $g_{\mu\nu}$ turn out to be
\begin{align}
{\cal H}_{00}=&-\Sigma \left(\beta \Sigma^2 +3\alpha \Sigma +3 \right)g_{00},\\
{\cal H}_{ii}=& - \left[\gamma \left(\beta \Sigma^2 +2\alpha \Sigma +1 \right)+\Sigma\left(\alpha\Sigma+2\right)\right]g_{ii},
\end{align}
along with that of tensor $s_{\mu\nu}(f)$ in the $f$-sector  given by
\begin{align}
s_{00}=&-\left(\gamma-1\right) \Sigma \left(\alpha_4 \Sigma^2 +3\alpha_3 \Sigma +3 \right)f_{00},\\
s_{ii}=& -\left(\Sigma-1\right) \left[\left(\alpha_4\gamma +\alpha_3\right)\Sigma^2 +2\left(\alpha_3 \gamma+1\right)\Sigma +\gamma\right]f_{ii}.
\end{align}

Similar to the five-dimensional case, we will focus on the solution making the graviton terms constant:
\begin{equation}
\gamma =\Sigma = C_0,
\end{equation}
where $C_0$ is a constant, whose equation will be figured out later.
Note that this solution corresponds the proportional metrics, i.e., 
\begin{equation}
f_{\mu\nu}=(1-C_0)^2 g_{\mu\nu}.
\end{equation}
Hence, the corresponding effective cosmological constants in $g$- and $f$-sectors can be evaluated as follows 
\begin{eqnarray}
\label{4d-physical-frw-lambda}
\Lambda_0^g &=& -\frac{C_0}{{\tilde M}_g^2} \left(\beta C_0^2 +3\alpha C_0 +3\right),\\
\label{4d-reference-frw-lambda}
\Lambda_0^f&=&\frac{C_0}{{\tilde M}_f^2 (1-C_0)^3} \left(\alpha_4 C_0^2 +3\alpha_3 C_0 +3\right).
\end{eqnarray}
As a result, the following equation of $C_0$ reads from an equality $\Lambda_0^g = (1-C_0)^2 \Lambda_0^f$ as
\begin{eqnarray} \label{4d-equation-of-C}
&&\beta C_0^3 +\left(\beta+3\alpha-\alpha_4 \tilde M^2  \right) C_0^2 -3\alpha_3 \left(\tilde M^2+1\right)C_0 \nonumber\\
&& -3\left(\tilde M^2+1\right) =0.
\end{eqnarray}
It turns out that in the dRGT limit, this equation will reduce to that derived in \cite{WFK}:
\begin{equation}
\alpha_4 C_0^2 +3\alpha_3 C_0 +3 =0.
\end{equation}
Once the value of $C_0$ is solved from the above equation, the effective cosmological constants $\Lambda_0^g$ and $\Lambda_0^f$ shown in Eqs. (\ref{4d-physical-frw-lambda}) and (\ref{4d-reference-frw-lambda}) will be calculated accordingly. It appears that these constants are consistent with that investigated in previous papers on the Schwarzschild solutions of massive bigravity, e.g., see Eqs. (20) and (23) in the first paper listed in \cite{blackholes-review}. Hence, it might be expected that the expressions in Eqs. (\ref{4d-physical-frw-lambda}) and (\ref{4d-reference-frw-lambda}) along with the corresponding equation (\ref{4d-equation-of-C}) might also be valid for other  metrics $f_{\mu\nu}$ proportional to $g_{\mu\nu}$ of the four-dimensional bigravity. 

It appears that $\Lambda_0^f$ for the $f$-sector as defined in Eq. (\ref{4d-reference-frw-lambda})  vanishes  when 
\begin{equation}
 \alpha_4 C_0^2 +3\alpha_3 C_0 +3 =0
\end{equation}
assuming that $C_0 \neq 1$. Solving this equation gives us non-trivial solutions of $C_0$:
\begin{equation} \label{C0-eq1}
C_0=\frac{-3 \alpha _3 \pm \sqrt{3\left( 3\alpha _3^2-4 \alpha _4\right)}}{2 \alpha _4},
\end{equation}
with the requirement that $\alpha_3^2 >(4/3) \alpha_4$. Note that the corresponding $\Lambda_0^g$ should also be zero. This implies that 
\begin{equation} \label{C0-eq2}
C_0 = -\frac{3}{\alpha_3}.
\end{equation}
 Equating Eq. (\ref{C0-eq1}) to Eq. (\ref{C0-eq2}) leads to a relation between $\alpha_3$ and $\alpha_4$:
\begin{equation}
\alpha_4 =\frac{2\alpha_3^2}{3}.
\end{equation}
As a result, these results are consistent with that investigated in the four-dimensional massive gravity \cite{WFK,TQD}, where the reference metric is assumed to be non-dynamical. Indeed,  the effective cosmological constant derived from the following massive graviton terms in the four-dimensional dRGT gravity will be zero if the $\alpha_4 =2\alpha_3^2/3$  \cite{WFK,TQD}.

Now, we will see whether the effective cosmological constants of five-dimensional massive bigravity reduce to that of four-dimensional one assuming they share the same proportional factor, i.e., $C_0={\cal C} $.  As a result,  it turns out that if the condition
\begin{eqnarray}
\label{condition1-reduce-to-4d}
\sigma {\cal C}^3 + 3\beta {\cal C}^2 +3\alpha {\cal C}+1 &=&0,
\end{eqnarray}
holds  then both $\Lambda_0^g$ and $\Lambda_0^f$ shown in Eqs. (\ref{4d-physical-frw-lambda}) and (\ref{4d-reference-frw-lambda}) will be recovered  from that defined  in the context of five-dimensional massive bigravity as shown  in Eqs. (\ref{L0}) and (\ref{LF}), respectively. 
\section{Conclusions}\label{con}
In this paper, we would like to study some higher dimensional scenarios of the  massive bigravity \cite{SFH,review-bigravity,higher-bigravity,higher-bigravity-more}. In particular, a five-dimensional massive bigravity model has been investigated systematically in this paper for some well-known metrics such as the FLRW, Bianchi type I, and Schwarzschild-Tangherlini ones, which have also been discussed in the context of five-dimensional dRGT gravity \cite{TQD}. Due to the dynamical feature of the reference metric $f_{\mu\nu}$ in the context of massive bigravity, its field equations turn out to be more complicated than that derived in the dRGT gravity, where the reference metric is assumed to be non-dynamical. In particular,  the general expression of the Einstein field equations of the reference metric has been derived along with that of the physical metric \cite{higher-bigravity,higher-bigravity-more}. Additionally, the following Bianchi identities for both physical and reference metrics have also been addressed consistently. As a result, the Bianchi constraints, especially ones for $f_{\mu\nu}$, have played an important role in order to simplify the Einstein field equations in both $g$- and $f$-sectors by making all graviton terms constant. 

It has appeared that the obtained solutions in the present paper are slightly different from that investigated in the five-dimensional massive gravity \cite{TQD} due to the dynamical property of the reference metric $f_{\mu\nu}$. In particular,  the field equations of reference metric in the dRGT theory are algebraic such that they can easily be solved. The corresponding graviton terms calculated from the physical and reference metrics, which are taken to be compatible with each other, are then automatically constant without introducing any further constraint. In the massive bigravity, the Bianchi identities have been applied for a number of physical and compatible reference metrics.  As a result, one of possible solutions satisfying the Bianchi constraints is that the reference metric is proportional to the physical metric, i.e., $f_{\mu\nu}=(1- {\cal C})^2 g_{\mu\nu}$ with ${\cal C}$ is a proportional factor \cite{higher-bigravity,higher-bigravity-more}.  Note that this solution has been shown to be valid for all  metrics studied in this paper. This result is also valid for the four-dimensional massive gravity studied in many published papers \cite{Hassan:2012wr,review-bigravity,blackholes-review}. Note that the equation determining the value of ${\cal C}$ in the five-dimensional bigravity has been defined in Eq. (\ref{eq-of-cal-C}).
Thanks to the constant-like behavior of massive graviton terms under the assumption that the refernce metrics are proportional to the physical metrics, we have been able to derive some cosmological solutions for both $g$- and $f$-sectors. We have also examined whether the effective cosmological constants derived from four-dimensional massive graviton terms can be recovered in the context of five-dimensional massive bigravity. 

Similar to the massive gravity \cite{TQD,WFK}, the stability analysis has  been performed for the Bianchi type I in order to test the validity of the no-hair conjecture proposed by Hawking and his colleagues long time ago \cite{Hawking,counter-example,WFK1}. As a result, we have shown that the obtained five-dimensional Bianchi type I metrics are indeed stable against field perturbations. This means that the cosmic no-hair conjecture is indeed violated not only in the massive gravity \cite{TQD,WFK,Gumrukcuoglu:2012aa} but also in the massive bigravity. For the stability of the Schwarzschild-Tangherlini-(A)dS black holes, it needs further investigations, which should follow the results analyzed in papers  \cite{blackholes-stability}  for the four-dimensional Schwarzschild black holes in the context of massive (bi)gravity. One should also be aware of the stability analysis done in papers \cite{5d-sch-stability}, which deal only with massless gravitons, for the Schwarzschild-Tangherlini-(A)dS black holes.

Besides the Schwarzschild-Tangherlini-(A)dS black holes, one could expect that other higher dimensional black holes \cite{5d-review} such as the Myers-Perry black holes \cite{5d-kerr} might also exist in the five-dimensional massive  bigravity since the Kerr black holes have been shown to appear in the four-dimensional massive bigravity \cite{blackholes}. One could also consider five- (or higher) dimensional scenarios for some interesting extensions of the massive bigravity such as the $f(R)$ bigravity  \cite{mod1}, the scalar-tensor bigravity  \cite{mod2}, and the massive bigravity with non-minimal coupling of matter \cite{non-minimal}. A higher dimensional version of the multi-metric gravity \cite{NK,Hassan:2012wr} should also be investigated in the near future.  Finally, we would like to note that  a detailed confirmation of the ghost-free property of higher dimensional massive (bi)gravity should be done by one way or the other, although this task might be straightforward as claimed in Ref. ~\cite{higher-bigravity} since one might follow the proofs for the four-dimensional massive (bi)gravity {}\cite{proof,review-bigravity}.

We hope that the present study along with that in \cite{TQD} could shed more light on the nature of massive (bi)gravity as well as its extensions.


\acknowledgments
The author would like to thank Prof. S.~F.~Hassan and Drs.  K.~Hinterbichler and E.~ Babichev  very much for their correspondence. The author highly appreciates Drs. A.~Schmidt-May, M.~von Strauss,  R.~Brito, and  M.~Crisostomi  for their fruitful comments. 
The author is deeply grateful to Prof. W.~F.~Kao of Institute of Physics in National Chiao Tung University  for his useful advice  on massive (bi)gravity. This research is supported in part by VNU University of Science, Vietnam National University, Hanoi.

\end{document}